\documentclass{aa}
\usepackage{txfonts}
\usepackage{apj2aa}
\usepackage{astron} 
\usepackage{float,epsfig,psfig}
\usepackage{pstricks}

\begin{document}

\title{The X-ray luminosity function of galaxies in the Coma cluster}
\author{A. Finoguenov\inst{1,5}, U.G. Briel\inst{1}, J. P. Henry\inst{2},
 G. Gavazzi\inst{3}, J. Iglesias-Paramo\inst{4}, A. Boselli\inst{4}}

\offprints{A. Finoguenov, alexis@xray.mpe.mpg.de}

\institute{Max-Planck-Institut f\"ur extraterrestrische Physik,
             Giessenbachstra\ss e, 85748 Garching, Germany 
\and
 Institute for Astronomy, University of Hawaii, 2680 Woodlawn Drive,
  Honolulu, Hawaii 96822, USA
\and
Universit\'a degli Studi di Milano-Bicocca, Piazza della Scienza 3,
20126 Milano, Italy
\and
Laboratoire d'Astrophysique de Marseille,
BP8, Traverse du Siphon, 13376 Marseille, France
\and
Space Research Institute, Profsoyuznaya 84/32, Moscow, 117810, Russia
}

\date{Received  2003, November 28; accepted 2004, February 10}
\authorrunning{Finoguenov et al.}
\titlerunning{XMM sources in the Coma cluster.}

\abstract{The XMM-Newton survey of the Coma cluster of galaxies covers an
area of 1.86 square degrees with a mosaic of 16 pointings and has a total
useful integration time of 400 ksec. Detected X-ray sources with extent less
than $10^{\prime\prime}$ were correlated with cataloged galaxies in the Coma
cluster region. The redshift information, which is abundant in this region
of the sky, allowed us to separate cluster members from background and
foreground galaxies. For the background sources, we recover a typical
$LogN-LogS$ in the flux range $10^{-15}-10^{-13}$ ergs s$^{-1}$ cm$^{-2}$ in
the 0.5--2.0 keV band. The X-ray emission from the cluster galaxies exhibits
X-ray colors typical of thermal emission. The luminosities of Coma galaxies
lie in the $10^{39}-10^{41}$ ergs/s interval in the 0.5-2.0 keV band. The
luminosity function of Coma galaxies reveals that their X-ray activity is
suppressed with respect to the field by a factor of 5.6, indicating a lower
level of X-ray emission for a given stellar mass.   }

\maketitle
\section{Introduction}

Studies of X-ray emission from normal galaxies can be used to reveal signs
of recent star-formation activity (Grimm et al. 2003), the presence of the
hot gas, filling the potential well of the galaxy and its immediate
surroundings (Forman, Jones, Tucker 1985), and the population of discrete
X-ray galactic sources (e.g. Pietsch et al. 1994, Irwin et al. 2003). With
the advent of high-spatial resolution observations, such studies become
feasible also for nearby clusters of galaxies, with the advantage that many
galaxies are observed simultaneously, all at essentially the same distance
from the observer. Such observations probe the effect of the cluster
environment on properties of galaxies, in particular ram-pressure stripping
and compression.

In this paper we present the properties of point sources in the XMM-Newton
mosaic of the Coma cluster of galaxies, which covers most of the cluster
virial radius. The advantages of studying the Coma cluster are the
availability of deep optical catalogs, plus the results of many studies at
various wavelengths, providing a unique opportunity for the identification
and in-depth study of the X-ray characteristics of cluster galaxies. We use
the catalog of Godwin, Metcalfe, Peach (1983, hereafter GMP) to associate
X-ray sources with galaxies and employ the
GOLDmine\footnote{http://goldmine.mib.infn.it/} (Gavazzi et al, 2003)
database to provide a wealth of additional information about the galaxies.

The paper is organized as follows, in \S\ref{s:data} we describe our
analysis of XMM-Newton data. The X-ray emission of Coma galaxies is analyzed
in \S\ref{s:glx}. In \S\ref{s:lf} we derive the X-ray luminosity function of
the Coma cluster galaxies. We adopt $H_{\rm o}=70$~km~s$^{-1}$~Mpc$^{-1}$
and $\Omega=1$, with a resulting $D_{L}=100$ Mpc. One degree corresponds to
1.67 Mpc.

\begin{figure*}
\includegraphics[width=17.cm]{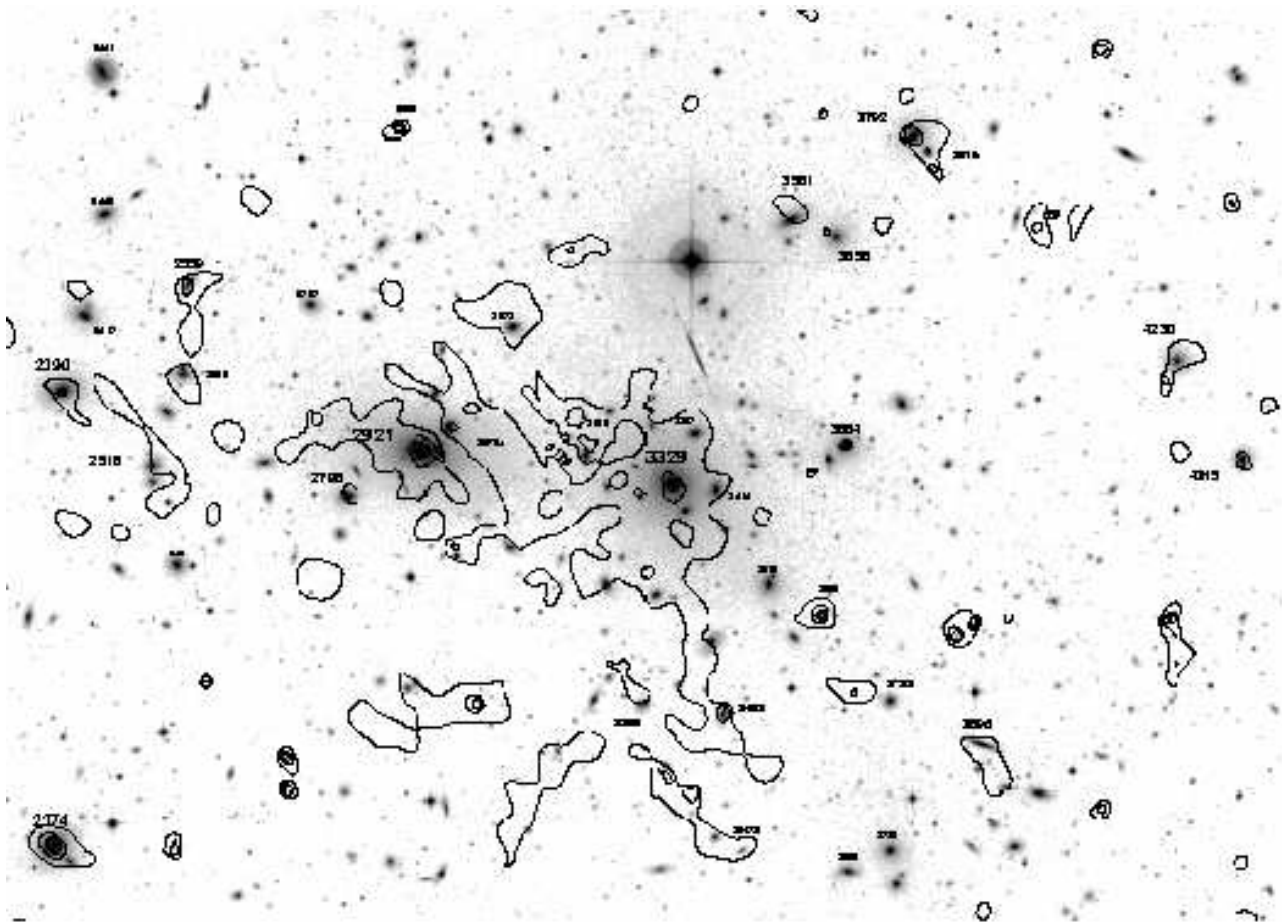}

\figcaption{DSS-2 B band image of the core of the Coma cluster (grey scale),
overlaid with EPIC-pn contours of the small-scale X-ray emission in the 
0.5-2.0 keV band.  Galaxies are labeled by their GMP catalog number. The image 
is centered on 194.913, 27.958 (J2000) and is 
$36^\prime \times 26^\prime$. North is up east is to the left.
\label{f:imh}
}
\end{figure*}

\begin{figure*}
\includegraphics[width=17.cm]{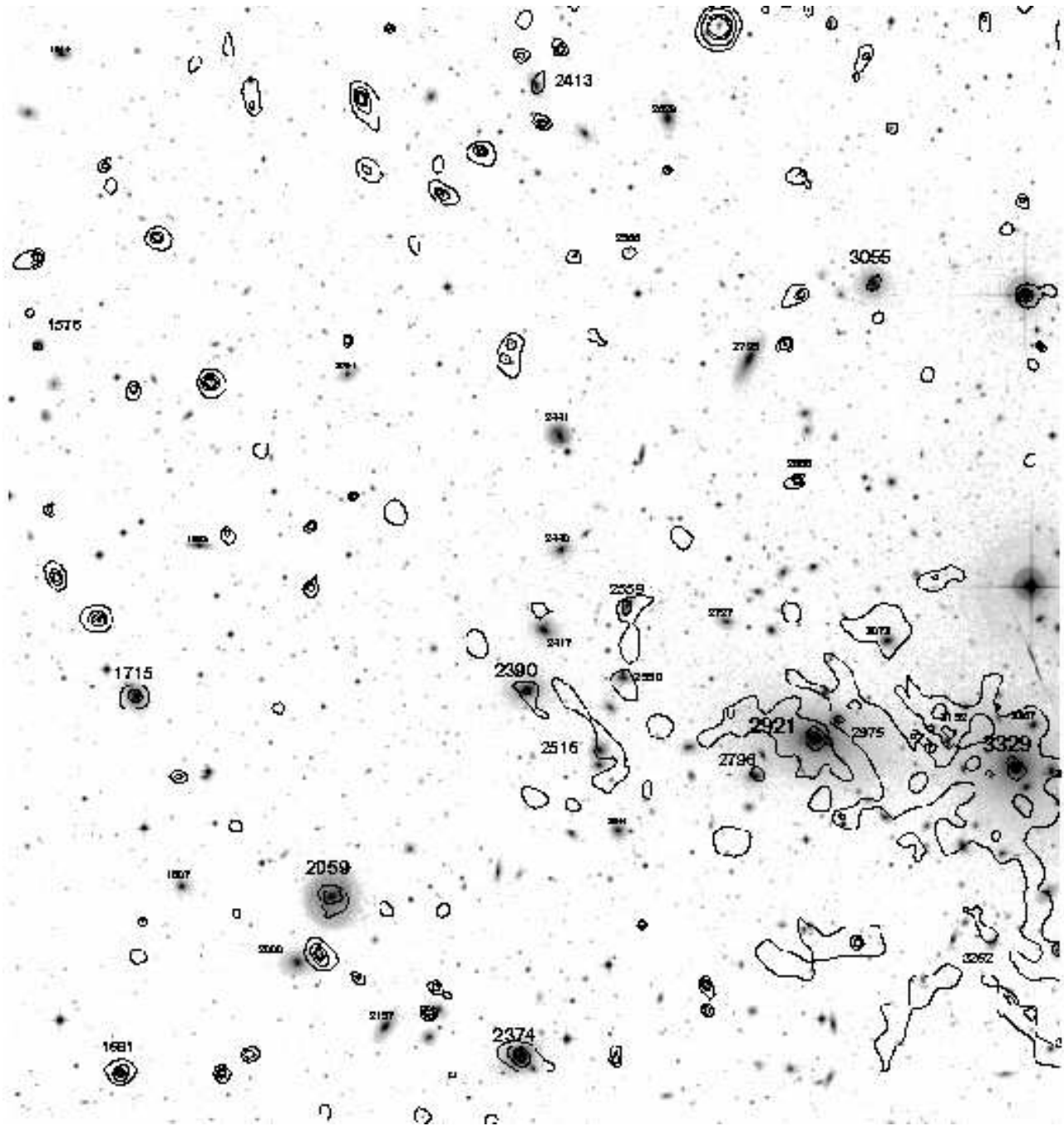}

\figcaption{Same as Fig.\ref{f:imh} for the north-east region of the Coma
cluster. The image is centered at 195.240, 28.095 (J2000) and is 
$37^\prime \times 39^\prime$. North is up east is to the left
\label{f:imhne}
}
\end{figure*}
\begin{figure*}
\includegraphics[width=17.cm]{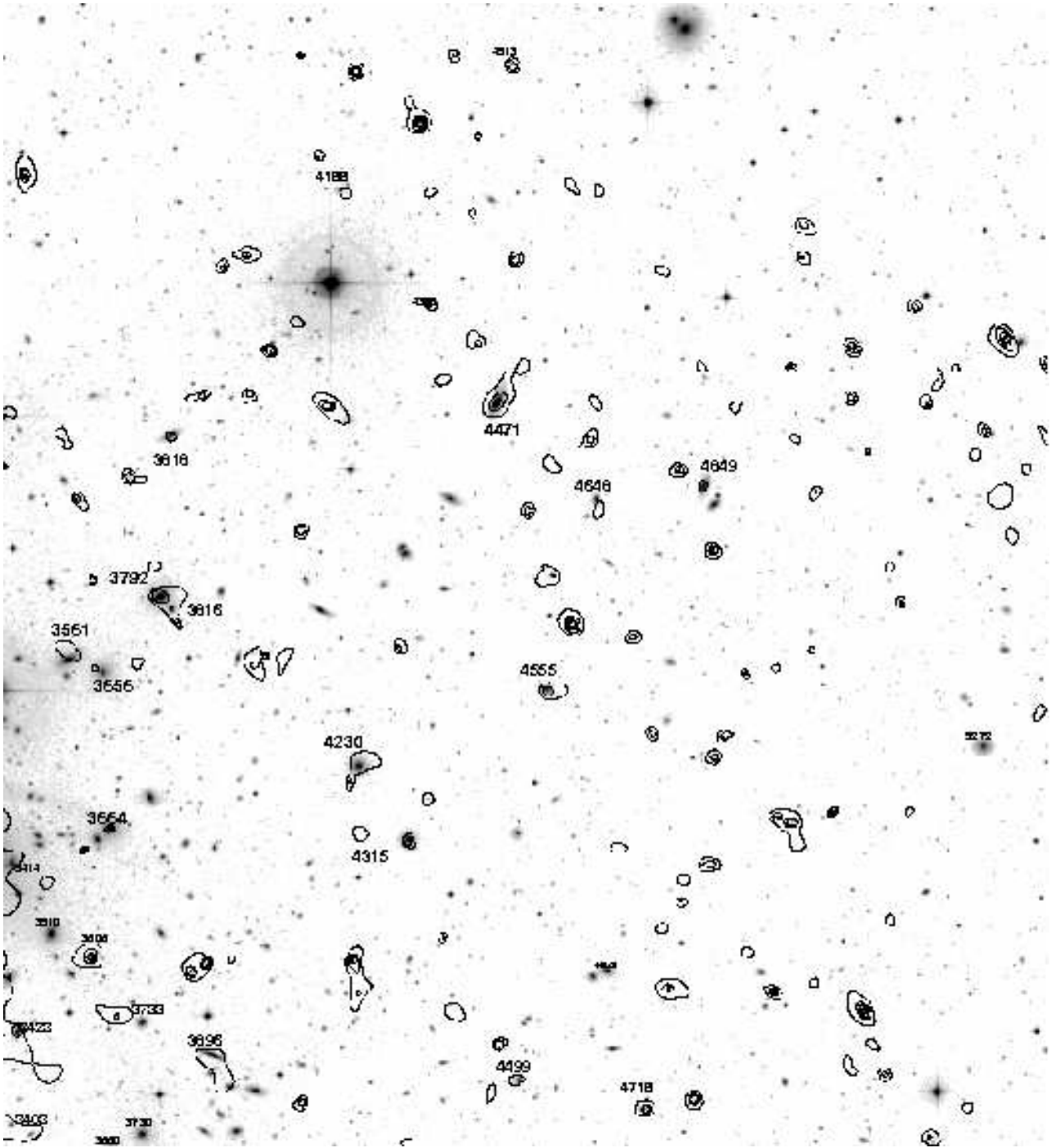}

\figcaption{Same as Fig.\ref{f:imh} for the north-west region of the Coma
cluster. The image is centered at 194.499, 28.134 (J2000) and is 
$39^\prime \times 44^\prime$. 
\label{f:imhnw}
}
\end{figure*}
\begin{figure*}
\includegraphics[width=17.cm]{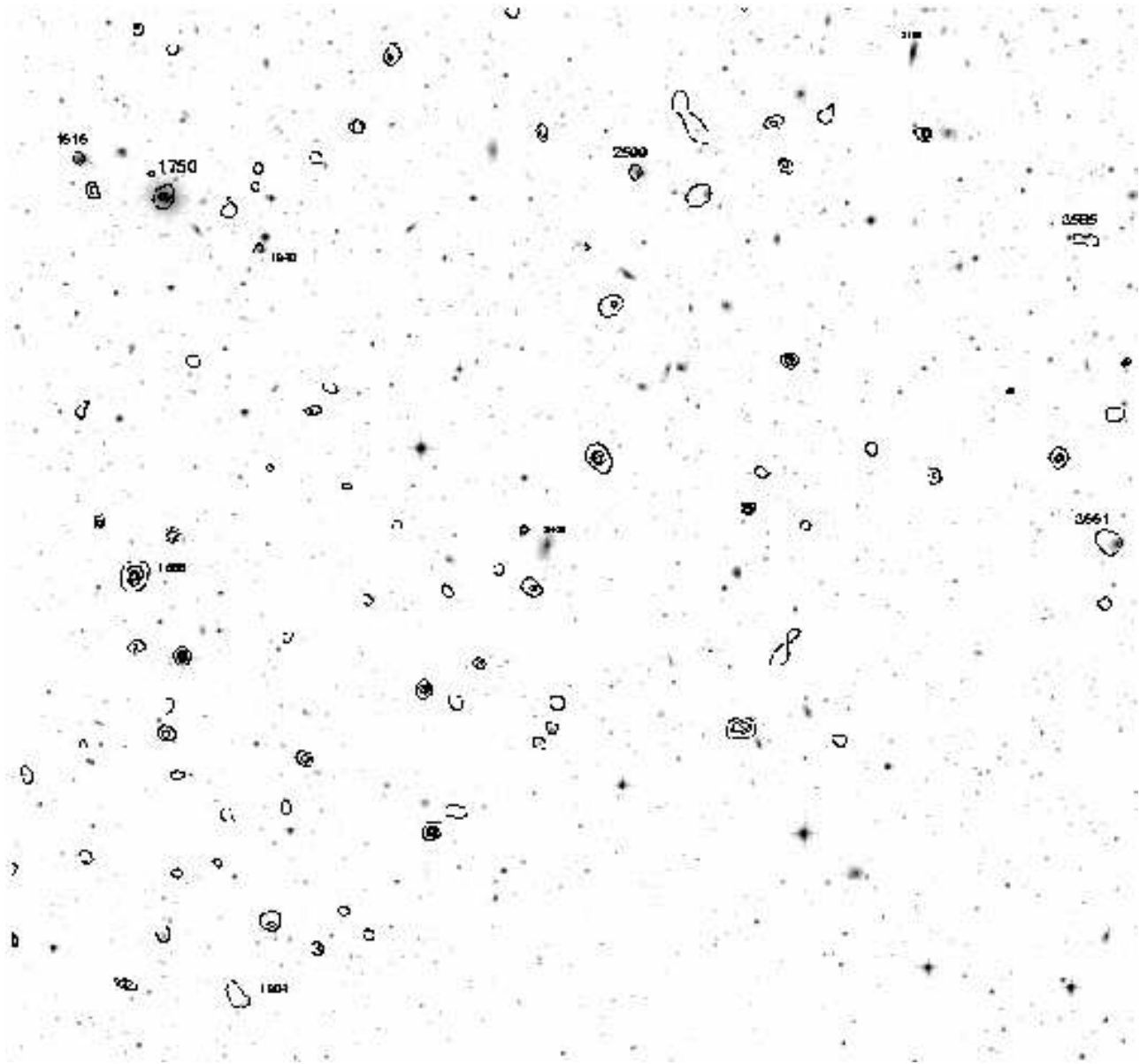}

\figcaption{Same as Fig.\ref{f:imh} for the south-east region of the Coma
cluster. The image is centered at 195.186, 27.409 (J2000) and is 
$43^\prime \times 40^\prime$. 
\label{f:imhse}
}
\end{figure*}
\begin{figure*}
\includegraphics[width=17.cm]{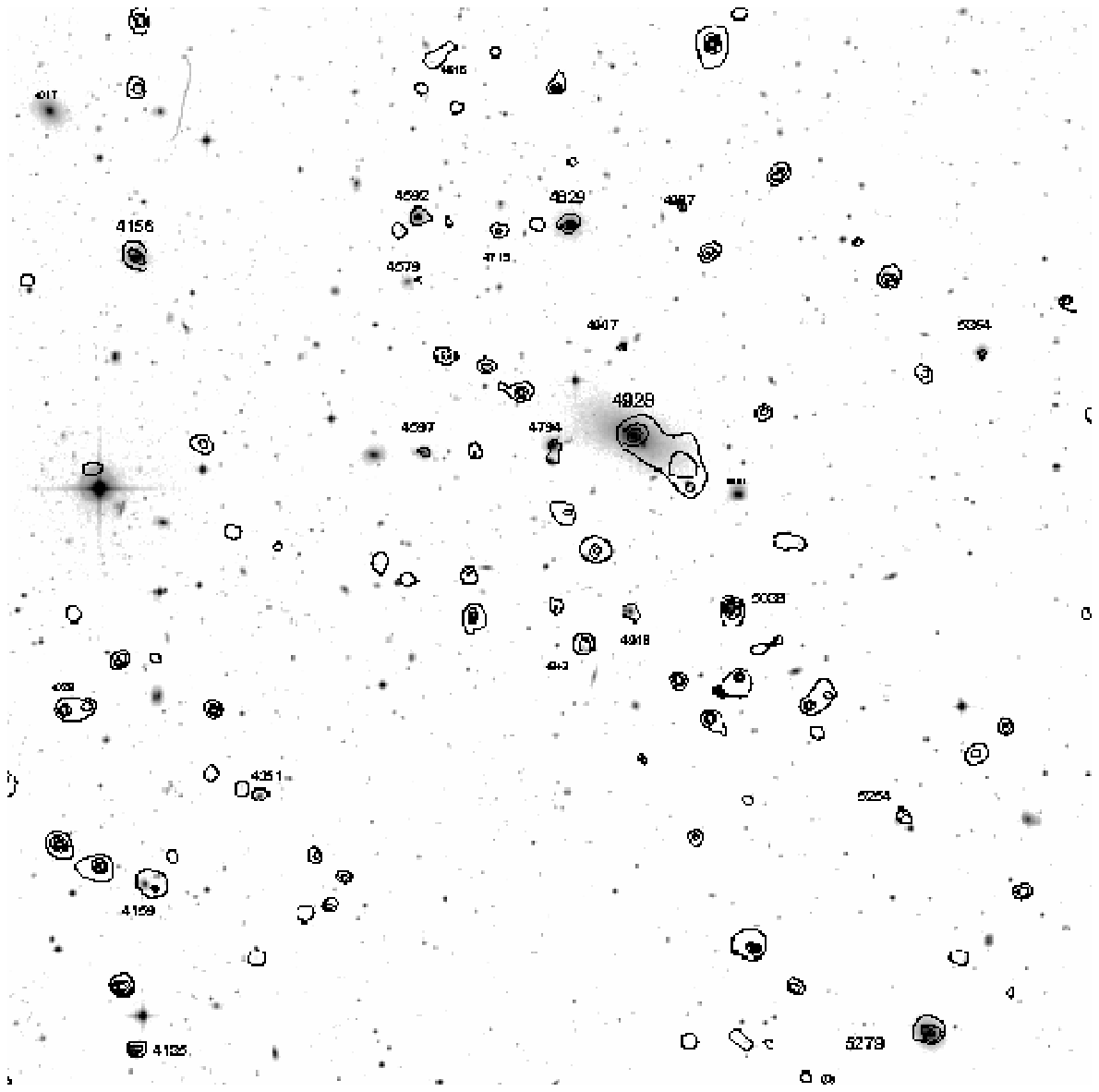}

\figcaption{Same as Fig.\ref{f:imh} for the south-west region of the Coma
cluster. There is one arcminute gap between this image and
Fig.\ref{f:imhnw}. The image is centered at 194.401, 27.440 (J2000) and is 
$35^\prime \times 34^\prime$. 
\label{f:imhsw}
}
\end{figure*}

\begin{figure*}
\includegraphics[height=18.cm,angle=-90]{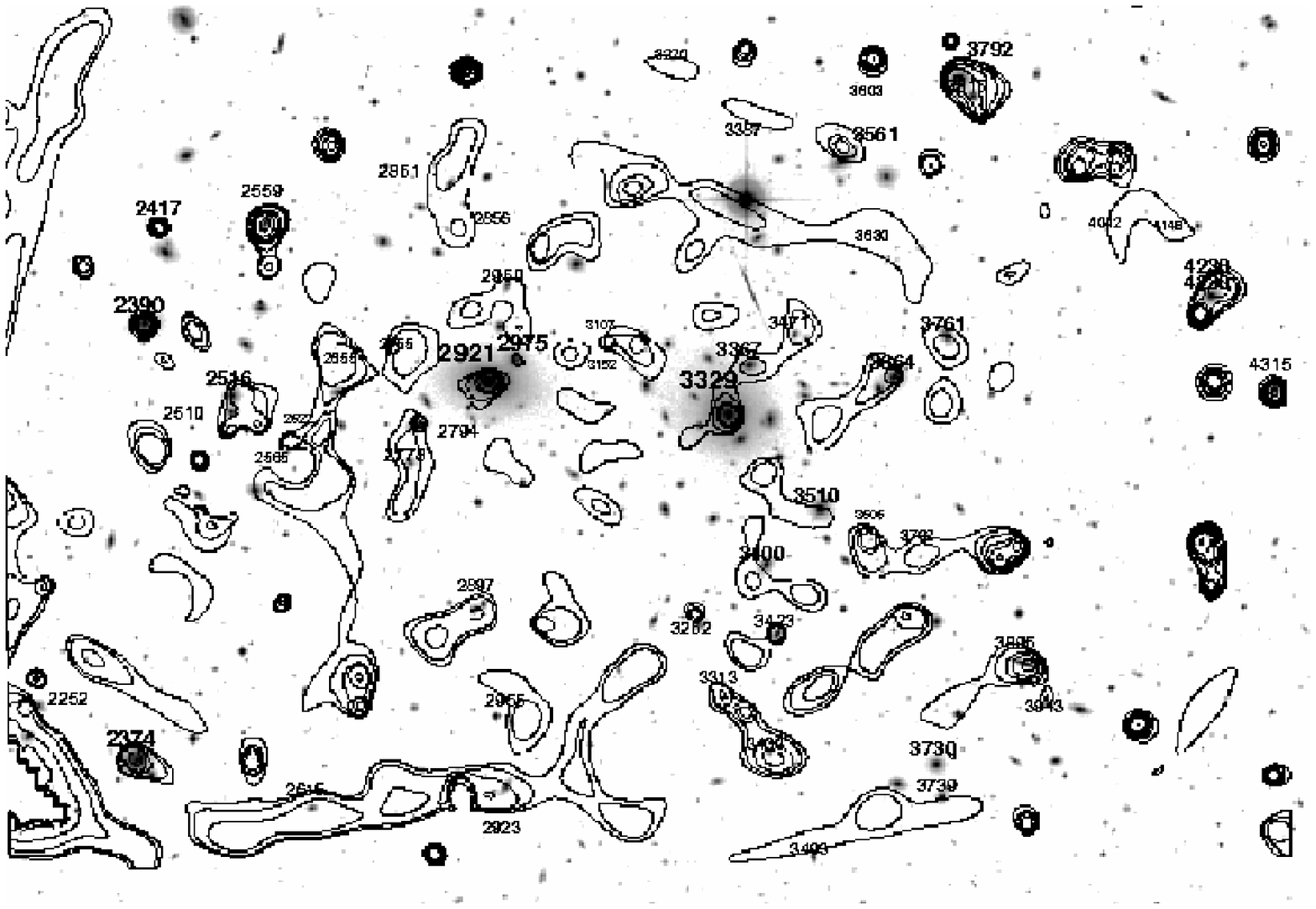}

\figcaption{DSS-2 B band image of the core of the Coma cluster center,
overlaid with the fluctuations in the entropy map of Coma cluster gas, as
observed by pn.  Galaxies are labeled by their GMP catalog number. The image 
is centered at 194.958, 27.939 (J2000) and is $40^\prime
\times 27^\prime$. North is up east is to the left
\label{f:imhent}
}
\end{figure*}

\begin{figure*}
\includegraphics[width=8.2cm]{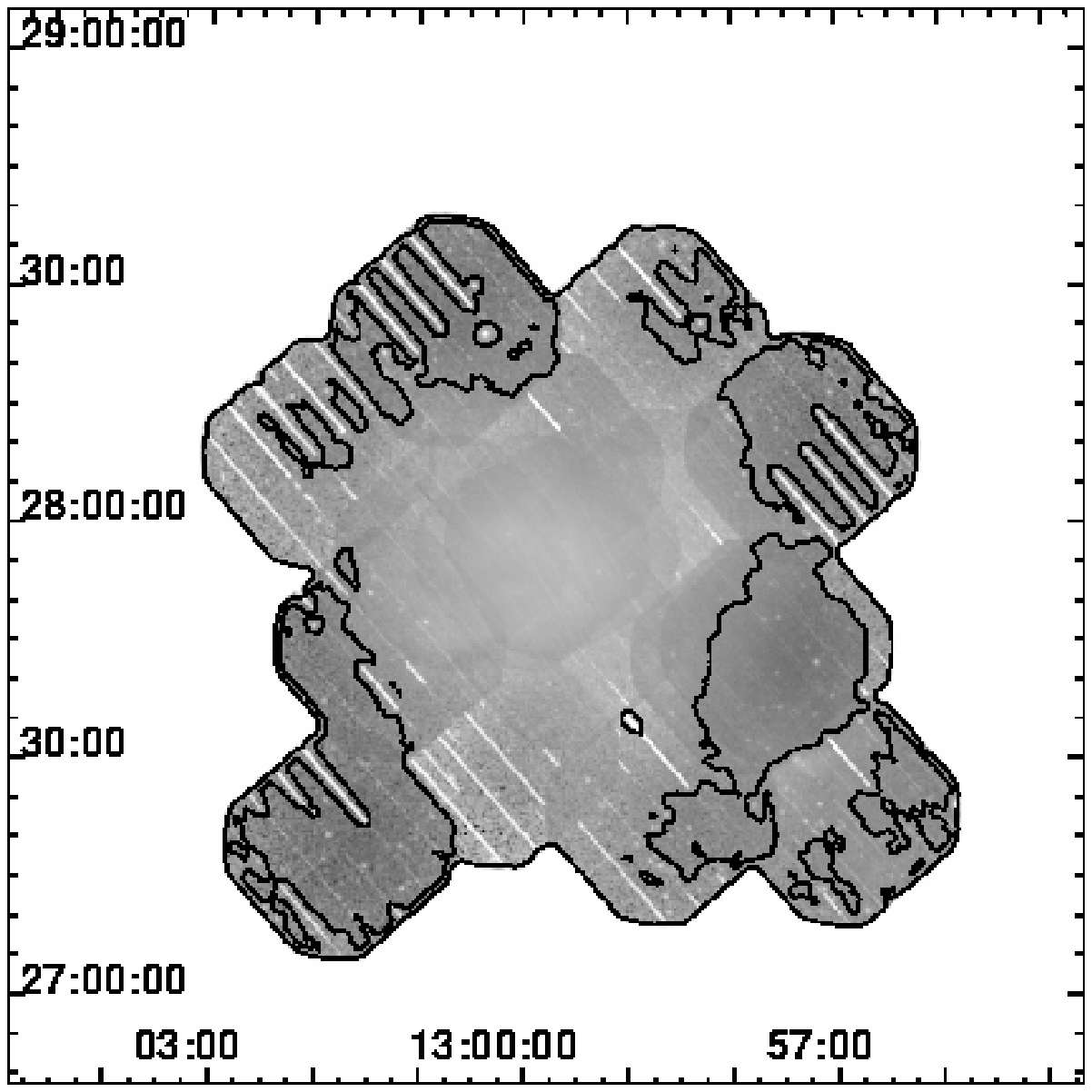}\hfill\includegraphics[width=8.2cm]{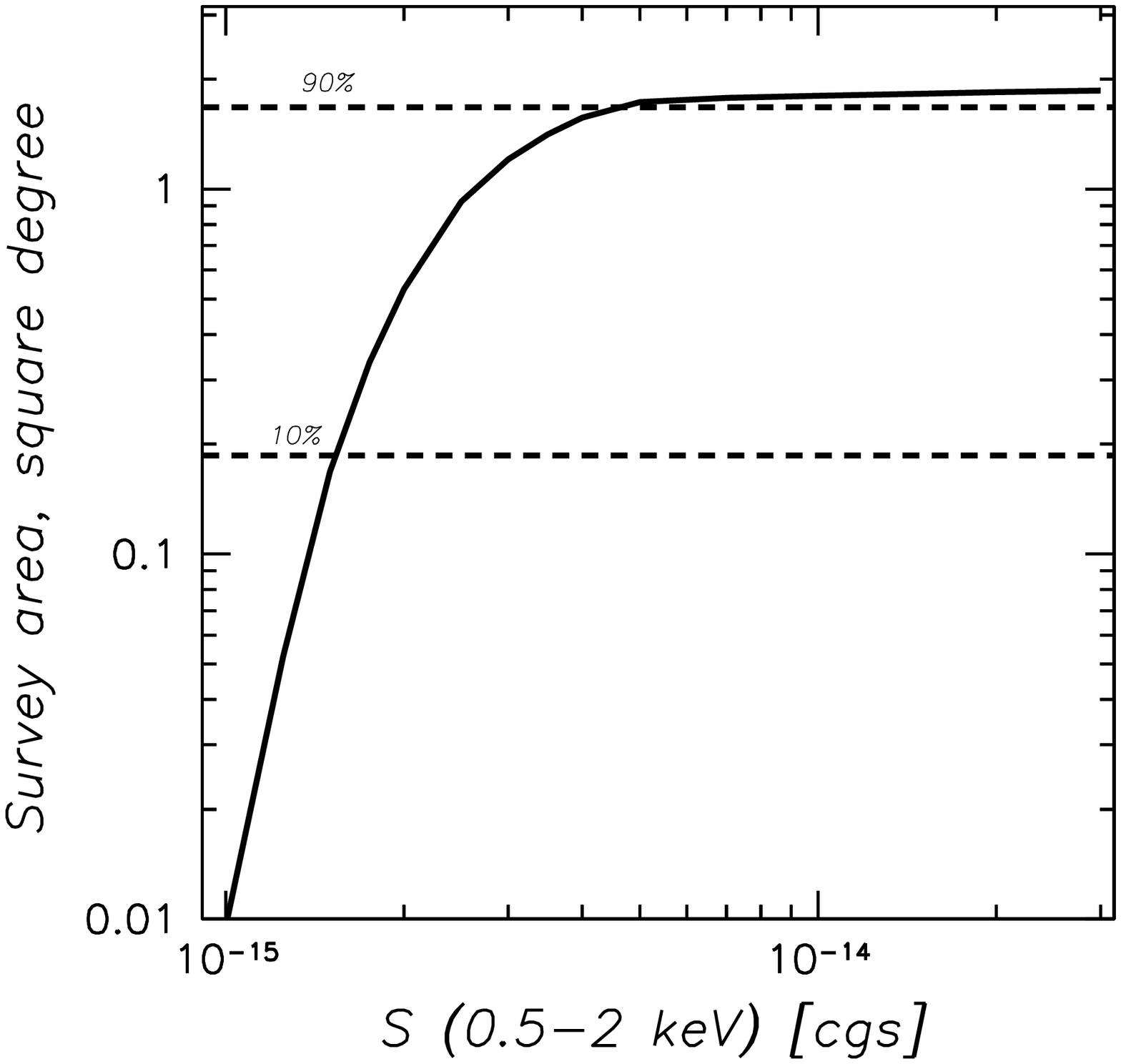}

\figcaption{{\it Left panel.}  Sensitivity map for source
detection on the $8^{\prime\prime}$(radius) wavelet scale. 
Contours enclose two areas of similar sensitivity, $10^{-14}$ 
ergs s$^{-1}$ cm$^{-2}$, which includes
the entire image, and $2\times10^{-15}$ ergs s$^{-1}$ cm$^{-2}$. {\it Right
panel.} Surveyed area vs. source flux in the $0.5-2.0$ keV band. Dashed
lines show at which flux the area reaches 10\% and 90\% of
the total area surveyed of 1.86 square degrees.
\label{f:area}
}
\end{figure*}

\section{X-ray observations and data reduction}\label{s:data}

The goal of this section is to describe our method of identifying an X-ray
source with a galaxy in the cluster. Here we provide illustrations that gave
helpful suggestions how such an identification should be defined. We attempt
to define our criteria rather loosely, but later sharpen them using the
experience with robust identifications. The selection criteria hopefully
will become transparent as we tabulate the results for intermediate
steps. In a brief, we remove X-ray sources from our catalog that are
identified with background AGN or central intracluster medium (ICM)
structures while retaining identifications with large separations between the
X-ray source and a cluster galaxy ascribing the separation to the effect of
ram-pressure stripping. For a subset of detected spiral galaxies, we show
that the position of the X-ray source corresponds to the position of the
stripped HI, thus reaffirming this later strategy.

In this paper we use the performance verification observations of the Coma
cluster obtained with the EPIC-pn instrument on board XMM-Newton (Jansen et
al. 2001). Preliminary reports of these observations were given by Briel et
al. (2001), Arnaud et al. (2001) and Neumann et al. (2001). In addition to
data reported in Briel et al. (2001), this work includes four more
observations: Coma-12 and Coma-13 fields, as well as another pointing on the
Coma center, performed to check the consistency of full frame and extended
full frame modes, and a re-observation of the Coma-2 field that was
contaminated by a high particle background in the original observation.

All observations have been reprocessed using the latest version of the XMM
reduction pipeline (XMMSAS 5.4), which yields astrometry to better than 1
arcsecond. A vignetting correction, crucial for obtaining reliable source
characteristics over a wide region, is performed using the latest
calibration (Lumb et al. 2003).

The images were extracted separately for each pointing, along with the
corresponding exposure maps. We select pn events with $PATTERN<5$ and $(FLAG
\& {\rm 0xc3b0809}) = 0$, which in addition to FLAG=0 events includes events
in the rows close to gaps and bad pixels, however it excludes the columns
with offset energy. This event selection results in a better spatial
coverage of the cluster, but at a somewhat compromised energy resolution,
which is sufficient for the broad-band imaging. When an X-ray photon
produces an electron cloud centered on a problematic region, such as a gap
or a bad pixel, most of the energy of the photon will be lost. Instead of
an event occupying two pixels (a double), we will detect an event occupying
one pixel (a single), but of much lower energy. 
We found experimentally that the above process was important in
the 0.2--0.4 keV band, producing bright columns near gaps. Given our
choice to include these columns in the image, we had to avoid energies
below 0.4 keV.

We employed the wavelet image reconstruction technique (Vikhlinin et
al. 1998) in order to begin separating the small-scale X-ray structure,
possibly associated with individual galaxies, from the large-scale structure
originating from the hot cluster gas. We set the wavelet peak detection
threshold to $4\sigma$, accepted flux down to $1.7\sigma$ and performed ten
iterations at wavelet scales of 4 and 8 arcseconds. The DSS2 B-band image of
Coma is overlaid in Figs.\ref{f:imh}-\ref{f:imhsw} with contours of the
X-ray emission detected on small scales using the 0.5--2.0 keV energy
band. Except for the Coma center, the identification of X-ray sources with
galaxies is unambiguous.  The nominal uncertainty in the astrometry of
XMM-Newton is $1^{\prime\prime}$. Our source detection method is subject to
an additional positional uncertainty of $2^{\prime\prime}$ (1/2 of the pn
pixel). These small errors make the identification of the point-like and
nearly point-like sources obvious.

Contrary to point sources, the identification of extended features is
non-trivial. In particular near the cluster center, the X-ray map exhibits
numerous extended fluctuations, which might be associated with individual
galaxies or be local enhancements of the ICM.  A strictly positional
criterion would cause us to reject the identification of those X-ray sources
that are slightly displaced from the optical galaxy position by the action
of ram-pressure stripping.  Thermodynamics helps us: since stripping implies
a gaseous origin for the offset emission, we expect it to have a typical
entropy of galactic gas, which is much lower than the entropy of the cluster
ICM.

The two giant galaxies at the center of Coma provide a good illustration of
the problem. Vikhlinin et al. (1994) using ROSAT data, found an emission
enhancement on scales of 1 arcminute. Using Chandra observations Vikhlinin
et a. (2001) showed that the gas associated with this scale has a
temperature of 10 keV, while there is a much more compact emission at $\sim
1$ keV temperature centered on both the galaxies. Based on the temperature,
Vikhlinin et al. (2001) concluded that the ROSAT detection is due to the
Coma ICM filling the potential wells of ellipticals. The hot gas is bound to
the potentials of these galaxies by the external pressure imposed by the
Coma ICM.  The difference between the Coma ICM and galaxy interstellar
medium (ISM) is revealed by their entropy. Although the origin of X-ray
temperatures in both ICM and ISM are shocks, the difference in the entropy
is due to a combination of different initial states and the strengths of the
shock, which increases with the potential of the system ($T\sim M/r$). In
the Coma cluster this difference is particularly large, so a separation
based on the entropy is fruitful.

We construct the entropy map using the wavelet-smoothed surface-brightness
map in the 0.8--2.0 keV energy range as an indicator of the electron density
squared and the hardness ratio map as the temperature distribution, and
defining the entropy as $S\sim T / \sqrt[3]{I}$. More details are given in
Briel et al. (2003). Valleys of low entropy, such as those presented in
Fig.\ref{f:imhent}, should enable us to locate X-ray sources associated with
the galaxy ISM. Enhancements of the emission due to ICM trapped in the
potentials of the galaxies due to the cluster overpressure should show up in
the gas pressure map, but not in the entropy map, as gas compression occurs
adiabatically. Also, the fluctuations that are associated with shocks in the
ICM, will (if at all) be seen as positive fluctuations in the entropy
map. The issue of ISM/ICM separation in not unique to identifying Coma
galaxies since Briel et al. (2001), among others, report a quite
inhomogeneous hardness ratio map of the Coma center, which is not
understood.

There additional uncertainties associated the low entropy ISM of galaxies
in a cluster. First it is subject to a number of instabilities, such as
cooling and a combination of stripping and Kelvin-Helmholtz instability and
should be understood as result of dynamical equilibrium between the
processes of gas removal and gas replenishment (e.g. Gaetz et al. 1987;
Matsushita 2001). However in Coma, an additional complication arises from
the presence of the large-scale low entropy gas associated with the
infalling group associated with the NGC 4911 galaxy, first
detected by White, Briel and Henry (1993) and further studied by Vikhlinin
et al. (1997) and Neumann et al. (2003). The group occupies the
south-eastern part of Fig.\ref{f:imhent}. A number of weaker large-scale
features are also evident in Fig.\ref{f:imhent}.

To summarize, we decided to use the following identification criterion: if a
negative valley in the entropy map or a positive peak in the emission map
lies within the size of the galaxy, it is considered to be associated with
the galaxy. The resulting detections of galaxies are presented in
Tab.\ref{t:x} and Tab.\ref{t:s} for the emission and entropy methods,
respectively. The entropy method is only reported for the center of the Coma
cluster where the level of the overall emission is high.

As discussed below, most of the point sources detected in the present survey
are identified with background AGNs. In order to compare with other AGN
studies, we selected the 0.5--2.0 and 2.0--4.5 keV bands. The flux is
extracted using a circle of $20^{\prime\prime}$ radius, centered on the
detected peak. The correction for flux outside these apertures is 1/0.74 in
the $0.5-2.0$ keV band and 1/0.73 in the $2.0-4.5$ keV band, using the
results of the in-flight calibrations of Ghizzardi (2001). Changes in the
PSF at large off-axis angles lead to variation of the aperture correction
within 1\% for both the soft and the hard band. The countrate-to-flux
conversion for $0.5-2.0$ keV band was done assuming a power law spectrum
with a photon index of 1.4, which for the medium filter used in this survey,
gives an effective area of the telescope of 1057 cm$^2$ and a mean energy of
1.03 keV.

For the sources identified with galaxies we convert the counts in the
$0.5-2.0$ keV band to the flux assuming a 1 keV thermal plasma of 0.3 solar
metallicity, yielding the effective area of the telescope of 1054 cm$^2$ and
a mean energy of the photon of 0.94 keV.

In estimating the net flux from the detected source, we also subtract the
background due to the diffuse emission of the Coma cluster. For this we use
a larger ($80^{\prime\prime}$ in radius) circle, centered on the source,
excluding the central $20^{\prime\prime}$, containing most of the source
flux (see above for details) and we scale the flux according to the relative
area--exposure--solid angle product (cm$^2$ s arcmin$^2$). This implicitly
assumes a flat distribution of the diffuse cluster X-ray emission on the
$80^{\prime\prime}$ scale, which is appropriate even for substructure in the
Coma diffuse emission. An underestimate of the source flux due to
subtraction of the scattered flux is less than 2\% in this
procedure. However, when halos of galaxies are filled with hotter medium,
like in the case of NGC4889 (GMP 2921), the use of the $80^{\prime\prime}$
aperture underestimates the 'background'.

We also tested the background subtraction in $40^{\prime\prime}$
aperture. Significant differences were found for galaxies GMP 2921, 3152,
3816.  For these galaxies a smaller aperture was adopted
($40^{\prime\prime}$ in radius) to estimate the Coma diffuse background with
resulting aperture correction of 0.68 due to the larger PSF scattered flux.

There is occasionally a difference in the reported flux for the same galaxy
between Tab.\ref{t:x} and Tab.\ref{t:s}, which is partly due to a different
center ascribed to the source. Also as removal of point sources is complete
only in the X-ray method, there is a possibility of an over-subtraction of
the background in the entropy method (Tab.\ref{t:s}). For most of the
sources, however, the difference is not very significant. Exceptions are GMP
2390, 3329, 3403.

The survey area is illustrated in Fig.\ref{f:area}, where we indicate the
location of zones where we reach higher sensitivity. Assuming a uniform
distribution of the surveyed sources, valid for background AGNs, in the
right panel of Fig.\ref{f:area} we show the one-dimensional effective area.
While a source brighter than $10^{-14}$ ergs s$^{-1}$ cm$^{-2}$ could be
detected anywhere in the totally surveyed 1.86 square degree area, only
within ten percent of the surveyed area it is possible to detect a source of
flux (here always in the 0.5--2.0 keV band) of $1.4\times10^{-15}$ ergs
s$^{-1}$ cm$^{-2}$, while 90 percent of the total surveyed area are already
available for sources with the flux of $4.8\times10^{-15}$ ergs s$^{-1}$
cm$^{-2}$.  The sensitivity of source detection is also a function of the
source extent and our calculation is only valid for sources with extent less
than $10^{\prime\prime}$.


\subsection{Identification of X-ray sources with Coma galaxies}

In order to produce a reliable estimate of the X-ray properties of galaxies
in the Coma cluster we first must assess the identification of the detected
X-ray enhancements with galaxies, then we must reject associations with
galaxies that are not members of the cluster.

As outlined above, we use an entropy map to check the most difficult region,
the Coma center. First, we list all the possible sources in
Tab.\ref{t:x}--\ref{t:s} and we select the ones identified with galaxies in
the GMP catalog (marked in Figs.\ref{f:imh}--\ref{f:imhent} with their GMP
catalog numbers). Second, we cross-correlate the sources found in the
surface brightness and entropy maps. A number of sources are detected in
both maps. Sources {\it not seen} in the entropy maps cannot be produced by
the thermal emission and were attributed to AGN. When such a source had a
large ($>4^{\prime\prime}$) X-ray to optical source separation, this must be
a chance coincidence with the background AGN and was therefore rejected from
the final list. Sources absent in the surface brightness, but present in the
entropy maps were considered to be possible associations with galaxy
emission (we mark them as S-sources in the final lists, with X-sources being
the sources identified only in the surface brightness map).

Robust X-ray source --- galaxy associations are those where the source is
within the effective radius of a galaxy brighter than $10^9
L_\odot$ in the B-band. For fainter galaxies we restrict the sample to 
identifications within $4^{\prime\prime}$, because we do not expect to
detect diffuse emission from such galaxies and the fraction of chance
identifications is high because of the large surface density of dwarf
galaxies.

The resulting list has been cross correlated with existing redshift
catalogs to assess cluster membership, defined as $4500<V_h<10000$. 
Cluster membership for the few remaining galaxies with $L_B<10^9 L_\odot$
and no redshift available was based on the color criterion:
$B-R<1.8$ (Poggianti et al. 2001).

\subsection{Comments on X-ray--galaxy associations not considered to be cluster members}
  
All sources that survive these cuts are considered to be Coma
cluster galaxies and we give some of their properties in Tables \ref{t:gm}
and \ref{t:gm2}. Below we comment on some of the excluded sources.

Sources absent in the entropy maps that are most likely
background AGNs are: GMP 3262, 3606, 3656. GMP 3269 is perhaps a
chance identification with the ICM structure in the Coma core.

GMP 2794 and 2798:  these galaxies are CGCG 160248a and b, which form an
interacting system. We formally list 2794 as the X-ray counterpart.

GMP 3220: very likely a chance identification, as the extent of the X-ray
source is much larger than the galaxy.

GMP 3606: UV detected by FOCA (Donas et al. 1995), thus probably a star
forming background object at 114990 km/s, also listed as a QSO in NED.

GMP 3702:  background object ($v>30000$ km/s)

GMP 2897, MRK 60, CGCG160-243: star forming pair with highly discrepant
redshifts.  The brightest one (CGCG160-243a) is detected in the UV by FOCA,
however due to the poor spatial resolution of FOCA an association with the
other member cannot be excluded. In addition, there is a third fainter
galaxy towards the South of CGCG160-243a, NGP9 F323-0986277 with a magnitude
(from NED) of 19.5.

As previously mentioned, all the confirmed sources have an X-ray
counterpart within one $r_e$. Thus, we classify the X-ray sources associated
with GMP 1688, 1904, 1940, 2566, 2888, 3262, 3269, 3585, 4028, 4358, 4513,
4579, 4615  as chance identifications.

Only three galaxies in the final list have no redshift information. Based on
color criterion outlined above, we tentatively assign GMP 2550 and 4718 as
members and 4845 as background.

\subsection{Comparison with HI observations}

We look for X-ray -- HI associations to further assess the reliability of
our method of identifying an X-ray source with a cluster galaxy even when
the positions do not agree at the $10^{\prime\prime}$ level.  As the
underlying reason for allowing such associations was stripping, similar
shifts should be observed in HI. Thus an alignment between the X-ray and HI
source could yield additional support for our method. We have used the
Bravo-Alfaro et al. (2001) catalog of spiral galaxies for the comparison.

First we describe non-detections and associations that do not appear to be
physically related. Some of spiral galaxies are outside our field of view,
so of course we do not detect them. We also do not detect the following
spiral galaxies at any wavelet scale: IC3913, Mrk 058, FOCA 0195,
KUG1258+287, all of which except for Mrk 058 (unless projection plays a role
here) are located at the periphery of the cluster.

There is a very extended ($1.2^{\prime}$ radius) faint structure
$22^{\prime\prime}$ away from the NGC 4907, which is a part of a chain of
three similar sources. We do not consider this association to be a cluster
identification due to the very different size of the X-ray source. We
tentatively assign such sources to debris of groups of galaxies, which
should not be counted as emission associated with individual galaxies.  In
addition to a number of filaments reported in Neumann et al. (2003), which
fall into this category, there is another new group of galaxies in both
optical and X-ray, with approximate position 194.71510 27.81647 (J2000).

Now we describe a number of apparent physical associations of X-ray sources
and spirals. KUG1255+275 (GMP 4351) is a weak X-ray source, centered on the
galaxy, so at the current sensitivity, there is no stripping observed at
X-ray, while it is observed in the HI data.  Mrk 057 (GMP 4135) has X-ray
emission that is offset to the North, in the same direction and magnitude as
one of two HI extensions.  In addition to the emission centered on NGC 4848
(GMP 4471, we observe a spectacular tail of X-ray emission to the
north-west, corresponding exactly to the HI. Intriguingly, HI is not
observed at the galaxy center at all.  In CGCG 160-086 (GMP 2599) the X-ray
source is offset to the east, as is the asymmetric part of the HI.  X-rays
are offset to the south-east of IC4040 (GMP 2559), exactly like the HI.  In
NGC 4911 (GMP 2374) the X-rays are produced by a strong point source
centered on the galaxy. However there is no detected X-ray emission
corresponding to the tail of diffuse HI emission to the south-west
coincident with a location of a smaller galaxy. For NGC4921 (GMP 2059) the
X-ray peak is at the position of the southern of the two HI blobs. The
X-rays are offset to the east of NGC4926-A (GMP 1616), similar to the HI,
but slightly displaced north of the HI peak still within the HI contours.

In general, we see a good deal of coincidence between the X-ray and HI and
at the same time some subtle differences that would be worth investigating
in detail.  This comparison lends support to our idea that slight offsets
between X-rays and optical images are due to stripping.

\subsection{Background sources}

The first step in background source removal was by considering the redshift
of every galaxy identified with an X-ray source. Next we turn to a
statistical background removal, as is often done in estimates of the galaxy
luminosity function in the optical. For the current survey we need to have
two probability functions: one the chance for a galaxy of a certain
magnitude to be a background object and two the chance for an object of a
given magnitude to emit X-rays at the flux level detectable in our survey.
The first estimate, based on the available redshift information for an
optically selected catalog of galaxies in the Coma cluster, complete to
$r=20$ magnitude, yields the following cumulative fractions of background
objects in bins of $r$ magnitude: 0 ($r<15$), 12\% (at $r\le16$), 26\% (at
$r\le17$), 65\% (at $r\le18$), 76\% (at $r\le19$), 85\% (at $r\le20$).

We estimate the probability for an {\it object} (excluding stars) of a given
r-band magnitude to be a bright X-ray source using the optical follow-up of
the Chandra Deep Field South (Giacconi et al. 2002). For $r\leq20$, typical
of Coma galaxies, this probability is $15\pm5$ \%. However, {\it all} these
objects are identified with QSOs or AGNs.  In fact in the Chandra Deep Field
South there are only 7 galaxies with $r<19$ and all of them have X-ray
fluxes an order of magnitude fainter than the limit of our survey. Within
the large uncertainty of this estimate, there is no contradiction between it
and our final list of background galaxies detected in X-rays at high fluxes.

The above agrees also with the more direct calculation of Georgakakis et
al. (2003), who have presented the contribution to the $logN-logS$ from
'normal' galaxy counts. Convolving their results with the sensitivity curve
of our survey, we expect 2.2 background galaxies to be detected in our
survey.

\includegraphics[width=8.4cm]{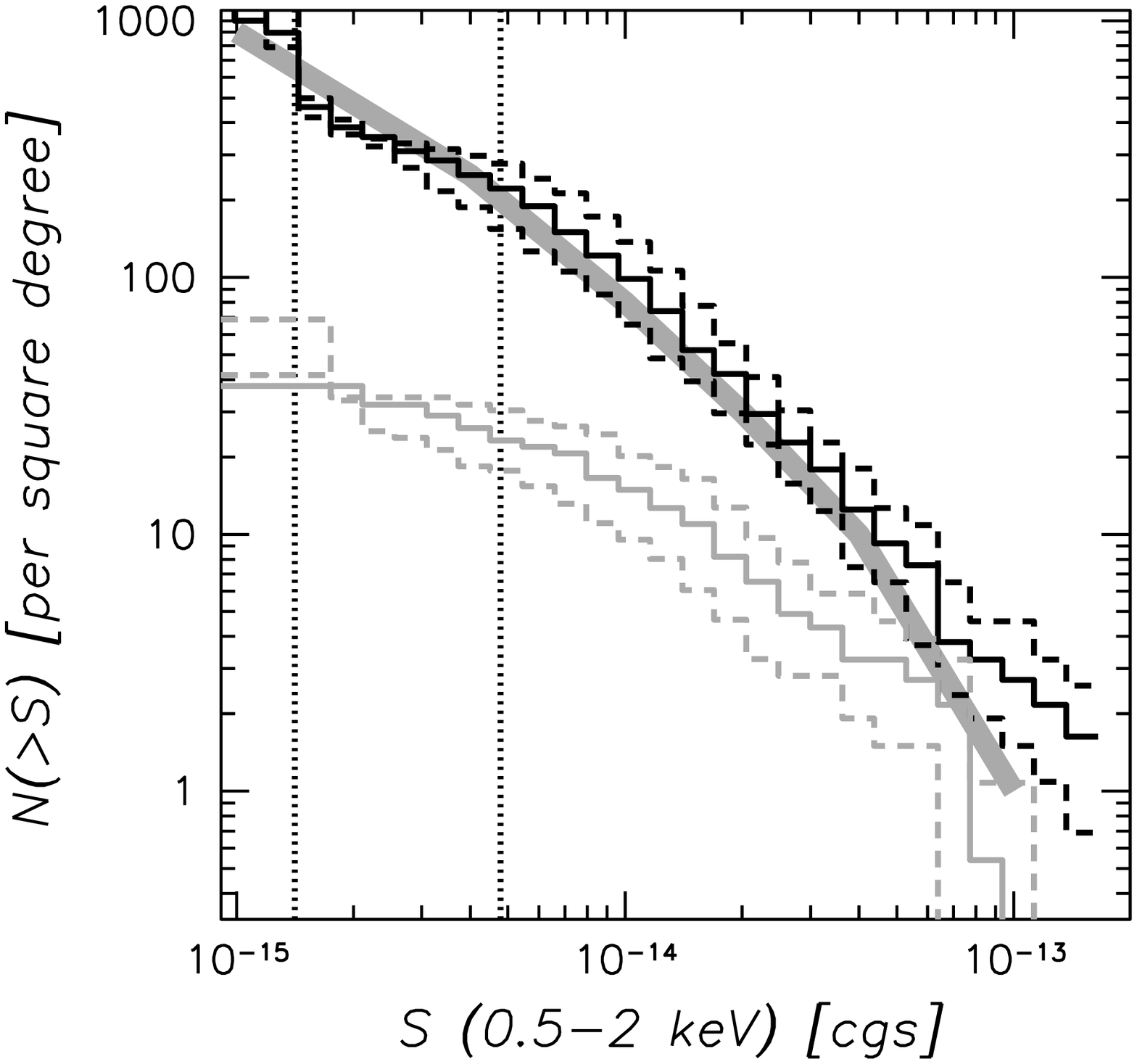}

\figcaption{$LogN-LogS$ of point sources in the Coma field. The black
histogram is the distribution of background objects, the grey histogram
shows the contribution at different fluxes of the removed cluster
galaxies. Dashed lines indicate uncertainties in the determination of the
$LogN-LogS$. The thick grey line is the AGN $LogN-LogS$ from Hasinger et al.
(2001). The vertical dotted lines show the flux at which the survey area
amounts to 10\% and 90\% of the total (1.86 square degree).
\label{f:logN}} 

Fig.\ref{f:logN} clearly shows that after removal of the sources identified
with member galaxies (as well as some obvious stars), our logN-logS relation
in the Coma field is consistent with previous estimates from ROSAT, XMM and
Chandra at a similar flux level.

\subsection{Comparison to other X-ray studies}

The first survey of X-ray emission of Coma galaxies was carried out by Dow
\& White (1995) using ROSAT PSPC data and a $1.5^\prime$ radius aperture to
extract the flux. Detection of a few galaxies was reported. For NGC 4889,
4874 and 4839 (GMP 2921, 3329, 4928) the galaxy flux is dominated by the
cluster component filling the potentials of these giant ellipticals, as has
already been shown by the analysis of XMM temperature structure and
therefore having much higher fluxes compared to our and Chandra (Vikhlinin
et al. 2001) results. In our analysis we do not consider ICM filling the
potentials of galaxies as a part of the X-ray emission from cluster
galaxies.

\includegraphics[width=8.4cm]{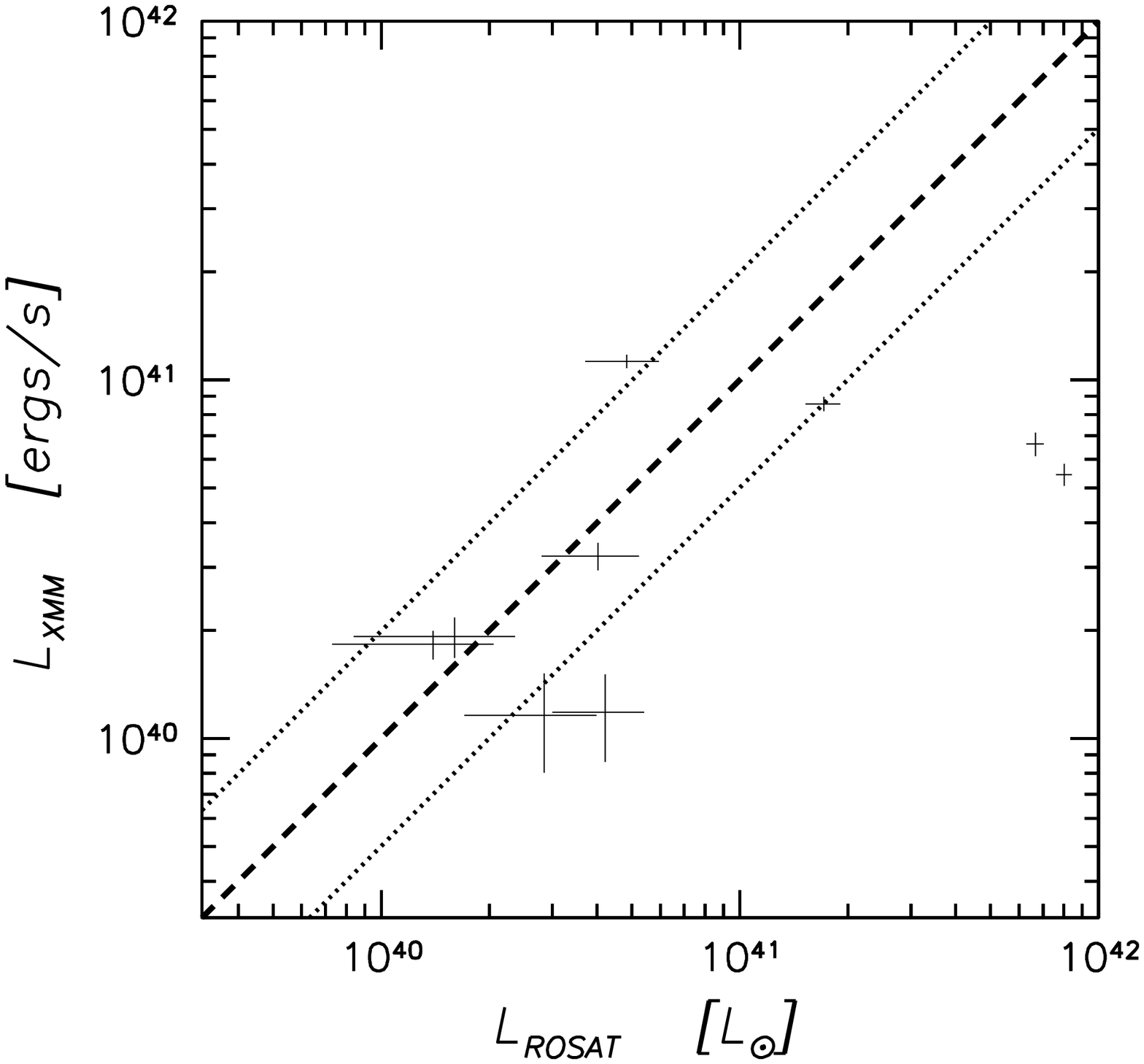}

\figcaption{Comparison of the 0.5--2.0 keV luminosity from XMM data with the
0.4--2.4 keV luminosity from the ROSAT data reported by Dow \& White
(1995). The slight difference in the energy bands can result in a 20\%
difference in the luminosities. The most outlying points are the two central
galaxies, NGC 4889 and 4874 (GMP 2921, 3329). The dashed line indicates
equal luminosities. Dotted lines indicate deviation by a factor of two. The
ROSAT values have been converted to our luminosity distance. The large
scatter between the results is attributed to the different apertures used
for the flux measurement.
\label{f:comp} }

In Fig.\ref{f:comp} we show a comparison of the luminosities derived
here with those reported in Dow \& White (1995). Apart from the two
central galaxies, the luminosities for the other seven sources are within a
factor of two of each other. The next most significant deviations are NGC
4839, another central galaxy mentioned above, and NGC 4911 (GMP 2374), the
only source were the XMM flux is significantly higher. As mentioned
previously, the origin of the X-ray emission from NGC 4011 is likely to be AGN
activity, for which variability by a factor of two is not surprising.

For some other galaxies, namely: NGC 4860, 4840, CGCG1256.1+2817, IC4040
(GMP 3792, 4829, 4230, 2559) the fluxes agree within the reported ROSAT PSPC
error bars. For NGC4898 (GMP 2794/2798), we find a factor of two fainter
source.  This source is close to the center of Coma, so the different spatial
resolution between ROSAT and XMM may cause the discrepancy.

We do not confirm the reported detections for NGC 4854 or IC 3959 (GMP 4017
or 3730). Our ($5\sigma$) upper limit on X-ray emission from NGC 4854 is a
factor of 3 lower than the $2.5\sigma$ detection of Dow \& White (1995). As
for IC 3959 we detect a point source identified with a faint optical
point-like object within the PSPC aperture. This object, being more than one
effective radius away from the galaxy ($45^{\prime\prime}$ compared to
$32.9^{\prime\prime}$), was dropped from the list. We identify this X-ray
source with a background quasar.  For the two central giant galaxies NGC
4889 and 4874 our flux estimate is consistent with Chandra (Vikhlinin et
al. 2001) if we take into account the somewhat larger aperture of XMM used
for our flux extraction.

\section{The nature of the X-ray emission from Coma cluster galaxies}\label{s:glx}

Although a number of X-ray detected galaxies exhibit star-formation
activity, as indicated by the UV (FOCA) observations of Donas et al. (1995),
no optical morphology peculiarities are seen on the DSS2 plates
(except possibly for dwarfs).

We note that a typical X-ray luminosity of galaxies in Coma does not exceed
$10^{41}$ ergs s$^{-1}$, implying that exotic sources associated for example
with extreme star-burst galaxies, such as the Antennae (Zezas et al. 2002)
are not present in our sample.  Luminosities up to $10^{41}$ ergs s$^{-1}$
are not unexpected even for the two dwarf galaxies we detect (GMP 2550,
4718), because these two reveal signatures of an ongoing merger.

\includegraphics[width=8.4cm]{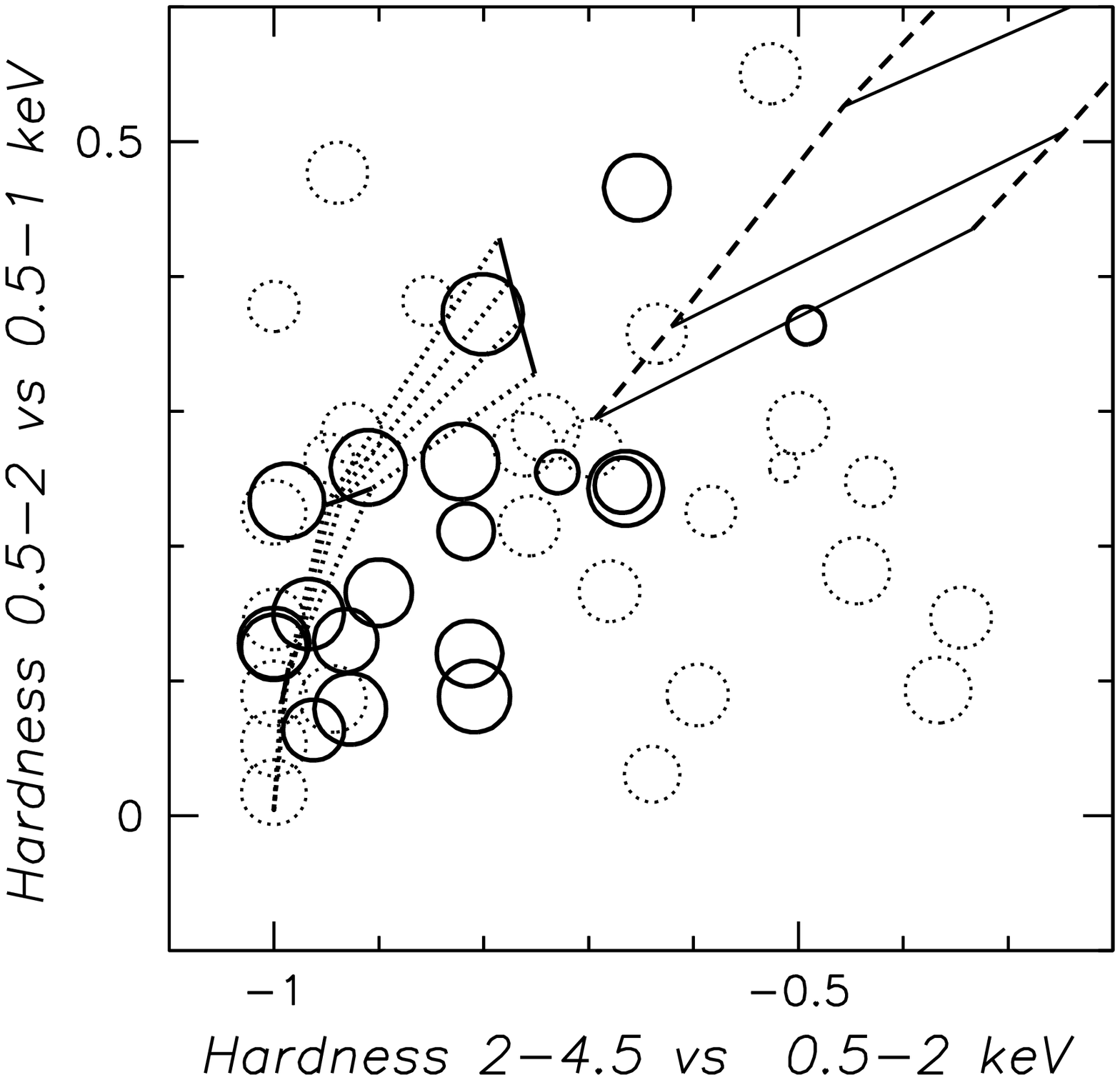}

\figcaption{X-ray spectral diagnostic diagram based on hardness ratios,
defined as (H-S)/H+S), where H and S corresponds to the counts in the hard
and soft energy bands. Data points are shown as circles with size
proportional to the absolute magnitude of the galaxy (solid circles mark
sources with an uncertainty of the hardness ratio $<0.15$, dashed circles
$<0.3$). The solid and dashed line grid gives the expected hardness ratios
for power law models with photon indices $\Gamma=1$ and 2 (left and right
dashed lines) and neutral hydrogen absorption of $logN_{\rm H}$ of galactic
value, 21, 21.5, 22 (bottom to top solid lines). The solid and dotted line grid 
gives the expectation for the emission of thermal gas with temperature between
0.1 and 2 keV (lower left to upper right solid lines) and element abundance
within 2 times solar (lower right to upper left dotted lines).
\label{f:hr-g}
}

\includegraphics[width=8.4cm]{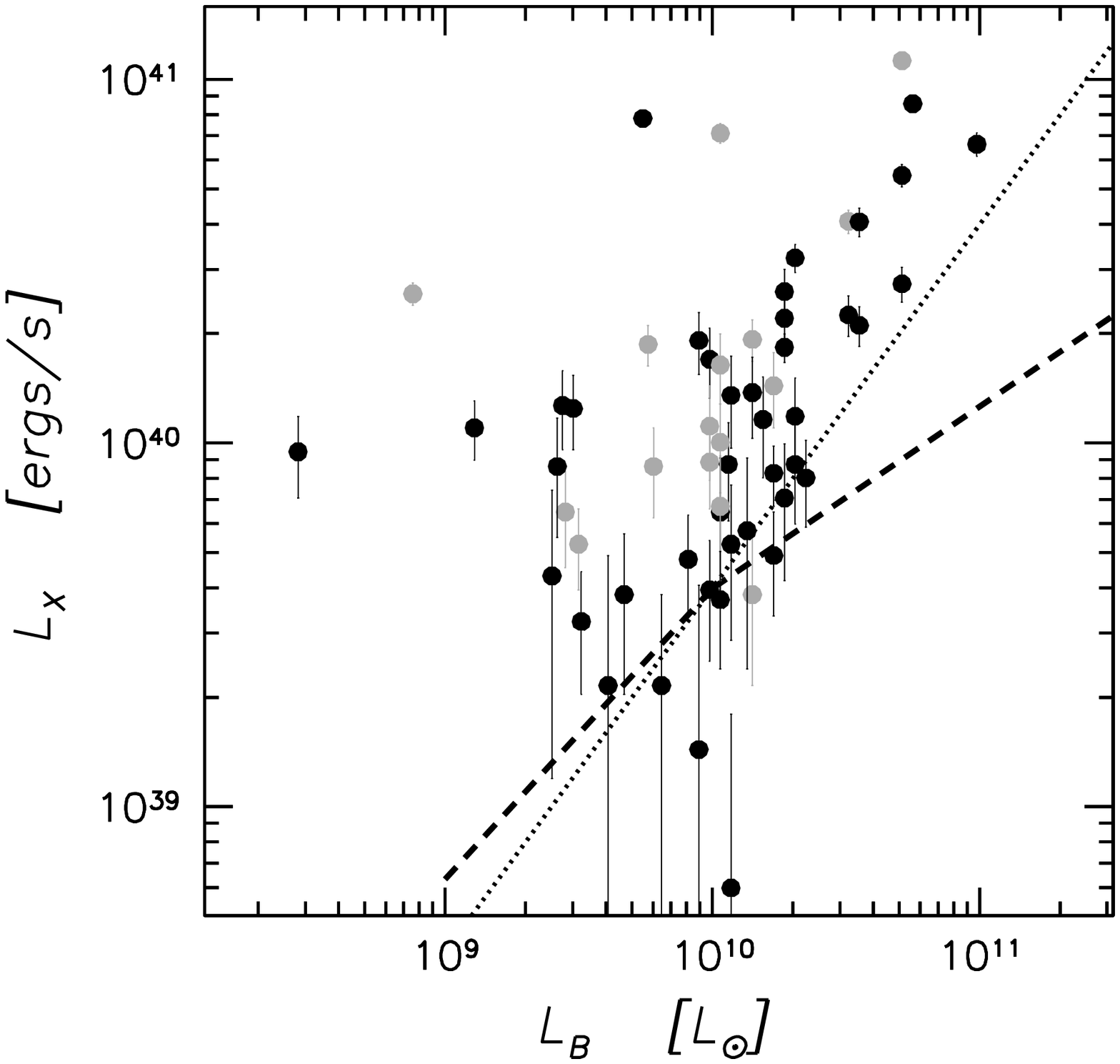}

\figcaption{ L$_{\rm X}-$L$_{\rm B}$ relation for sources identified with
Coma galaxies. The dotted line indicates the expected contribution from 
discrete galactic sources, scaled to the Chandra results on M84 (Finoguenov 
\& Jones 2001). The long dashed line is the expected discrete source
contribution using the apertures employed in the present survey The aperture 
correction is important even if flux were collected out to the effective radius, 
since the effective radius only contains half of the light. The grey points
indicate the galaxies with $UV-B<1$, according to Donas et al. (1995; FOCA telescope).
\label{f:glx}
}


To address the question of the origin of the X-ray emission from Coma
galaxies, we study their X-ray colors in Fig. \ref{f:hr-g}. The two grids
given in the figure allow us to disentangle the role of the diffuse
emission from the AGN activity in determining the X-ray luminosity.
Optically luminous gas-poor galaxies, whose emission is dominated by
integrated flux from point sources (LMXB) are expected to be found
within the power-law grid, as this emission is characterized by a
power law of photon index 1.4 and galactic nH (e.g. Finoguenov \&
Jones 2001). From Fig.\ref{f:hr-g} it is clear that most of the
sources have quite soft spectra, indicating a thermal origin from
diffuse gas. Some of the sources show spectral hardening,
possibly due to non-negligible contribution from an unresolved
population of LMXB.

\subsection{L$_{\bf X}-$L$_{\bf B}$ relation}

The L$_{X}-$L$_{B}$ diagram is a fundamental diagnostic tool for assessing 
the nature of the X-ray emission from Coma galaxies compared to galaxies 
detected in other surveys of the local universe with better spatial and 
spectral resolution. Fig.\ref{f:glx} presents such a diagram.  Approximately 14
objects are consistent with all the emission coming from discrete point sources,
when corrected for aperture effects.  This correction takes into account that
only a small fraction of the galaxy is observed in the present survey. Inclusion of
a larger fraction would lead to an overestimate of the flux due to the high
level of X-ray emission from the Coma ICM, which determines the effective
background of our X-ray data.  For the remaining two-thirds of the sample, the
contribution from discrete point sources is negligible.  To isolate
possible star-bursting objects, which are known to have their X-ray
luminosity significantly enhanced, we separate the sample according the UV-B
color. 
There is one detected dwarf star-forming galaxy
with an $L_X$ to $L_B$ ratio similar to the local galaxy Holmberg II (see
Zezas et al. 1999).

\begin{figure*}
\includegraphics[width=8.4cm]{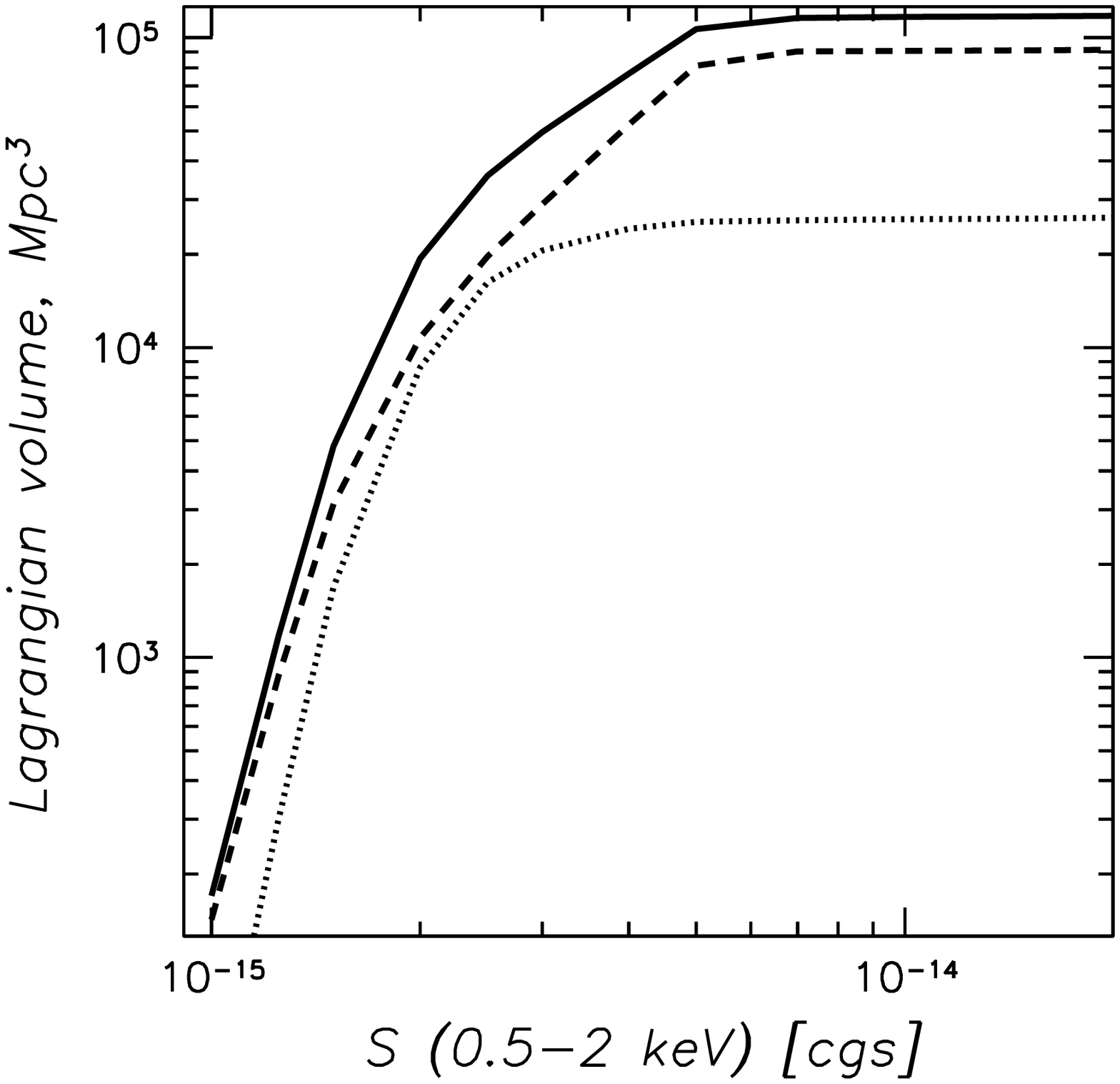}\hfill\includegraphics[width=8.4cm]{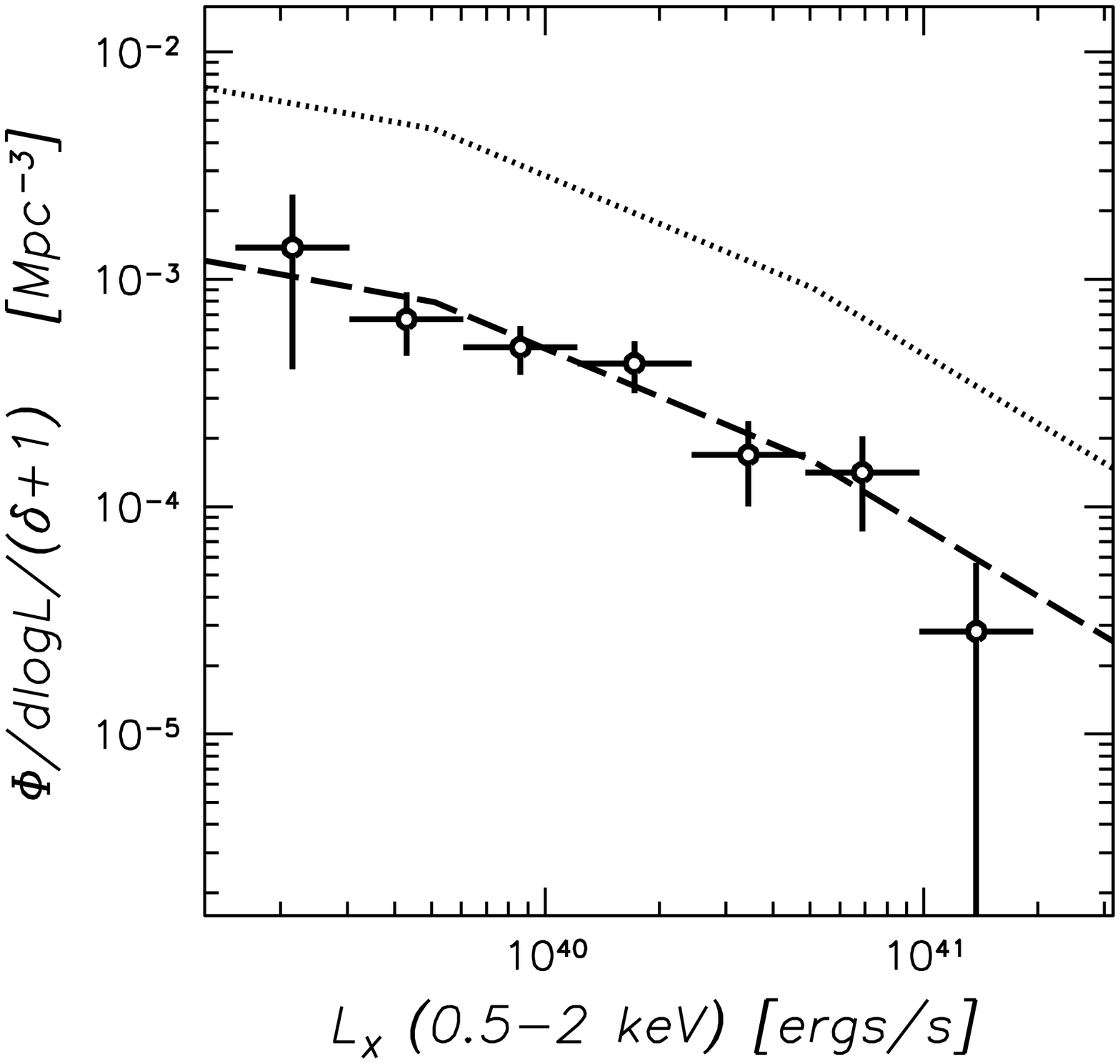}

\figcaption{{\it Left panel}. Lagrangian volume of the XMM-Newton survey of
the Coma cluster as a function of the flux in the 0.5-2 keV band. Solid line
indicates the total volume, while the dashed and dotted lines indicate the
contributions from the main and NGC 4839 sub-cluster.  {\it Right
panel}. the X-ray luminosity function of Coma galaxies. The dotted line
represents the X-ray luminosity function of local galaxies (Hasinger 1998)
scaled for the local overdensity and the dashed line represents the best fit
to the luminosity function of Coma galaxies.
\label{f:glf}
}  

\end{figure*}

\section{The X-ray luminosity function of Coma galaxies}\label{s:lf}

To determine the Coma galaxy luminosity function, we divide the
number of detected galaxies by the Lagrangian volume ($M_{\rm
surveyed}/\Omega_m/\rho_{\rm crit}$, where $M$ is the total gravitational
mass of the cluster) surveyed as a function of the flux.  We perform volume
estimates for the main cluster and the infalling NGC 4839 subcluster
separately applying the NFW dark-matter profile, $\rho\sim r^{-2.4}$ and the
corresponding virial ($r_{100}$) radius estimate of 4.0 and 2.6 Mpc (for
$\Lambda$CDM, Pierpaoli et al. 2001), using the $r_{500}-T$ relation of
Finoguenov et al. (2001).  The resulting Lagrangian volume is plotted in the
left panel of Fig.\ref{f:glf}.  As it can be seen from the figure, it is
important to account for the substructure, due to inhomogeneous sensitivity
of the survey.

The estimate of the total Lagrangian volume of the main cluster is $4 
\pi /3 \times\delta_{vir} \times r_{vir}^3=103060$ Mpc$^3$, where
$\delta_{vir}=100/\Omega_m=370$, so we survey 90\% of the virial mass.  The
mean radius of the survey in the observer's plane is 1.3 Mpc. A sphere of
this radius is characterized by a Lagrangian volume of 52000 Mpc$^3$, 57\%
of the survey volume.

The local luminosity function of X-ray sources by Hasinger (1998), scaled to
our assumption of $H_{\rm o}=70$~km~s$^{-1}$~Mpc$^{-1}$ is given in the
right panel of Fig.\ref{f:glf}.

The results of the CfA redshift survey (Santiago \& Strauss 1992) show that
the mean density of the typical volume covered by the local luminosity
function of Hasinger (1998) is three times the mean density of the
Universe. To provide a valid comparison, we further scale down the
luminosity function of Hasinger (1998) by a factor of 3 since we normalize
our Coma galaxy luminosity function to the mean density of the Universe.

We conclude that the X-ray emission of galaxies in the Coma cluster is
quenched on average by a factor of 5.6. We estimate this factor from the
difference between the luminosity function in the field and in the Coma
cluster. It seems natural to attribute this result to reduced star-formation
activity in cluster galaxies. Finoguenov \& Miniati (2004) show the
ellipticals in Coma have X-ray emission typical of the early-type galaxies
found outside the cluster environment. Thus, differences in the luminosity
functions between the Coma cluster and the field are due to preponderance of
cluster galaxies with low X-ray to optical flux ratios.

\begin{acknowledgements}

The paper is based on observations obtained with XMM-Newton, an ESA science
mission with instruments and contributions directly funded by ESA Member
States and the USA (NASA). The XMM-Newton project is supported by the
Bundesministerium f\"{u}r Bildung und Forschung/Deutsches Zentrum f\"{u}r
Luft- und Raumfahrt (BMFT/DLR), the Max-Planck Society and the
Heidenhain-Stiftung, and also by PPARC, CEA, CNES, and ASI.  The authors
thank the referee, Bill Forman, for useful comments, which improved the
presentation of the material of this work. AF acknowledges support from
BMBF/DLR under grant 50 OR 0207 and MPG.  AF profited from discussions with
Guenther Hasinger, Wolfgang Pietsch, Gyula Szokoly, Bianca Poggianti,
Michael Freyberg and Alexey Vikhlinin. USNO, DSS2, ADS and NED services are
acknowledged.
\end{acknowledgements}

\bibliographystyle{aabib99}

\begin{table*}[ht]
{
\begin{center}
\footnotesize
{\renewcommand{\arraystretch}{0.6}\renewcommand{\tabcolsep}{0.12cm}
\caption{\footnotesize
X-ray detections of galaxies towards the Coma cluster.
\label{t:x}}
\begin{tabular}{cccccccccccccc}
 \hline
X  & $lg(L_b)$ & B-R &  $r_e$ & GMP & separ.& \multicolumn{2}{c}{X-ray source
  J2000} & $F_{0.5-2}$ $10^{-15}$ & \multicolumn{4}{c}{$10^{-3}$ counts
  s$^{-1}$, band (keV):}\\
numb. & $L_\odot$ & &$^{\prime\prime}$ & & $^{\prime\prime}$&\multicolumn{2}{c}{RA Decl.} &ergs s$^{-1}$ cm$^{-2}$&0.5--2&0.5--1 & 1--2& 2-4.5\\
\hline
 19& 9.88& --& 15.9& 1576& 11.25& 195.5540& 28.2142&$ 7.4\pm1.9$&$ 5.2\pm1.3$&$ 2.9\pm1.0$&$ 2.2\pm0.9$&$ 0.2\pm 0.5$  \\
 25& 10.04& --& 22.4& 1616& 16.31& 195.5345& 27.6476 &$ 3.2\pm1.4$&$ 2.2\pm1.0$&$ 3.1\pm0.9$&$-0.9\pm0.3$&$ 0.4\pm 0.5$ \\
 33& 9.84& --& 12.8& 1681& 8.92& 195.5013& 27.7822 &$59.4\pm3.7$&$41.7\pm2.6$&$36.7\pm2.4$&$ 5.1\pm1.0$&$ 0.8\pm 0.5$ \\
 35& 8.77& 2.21& 8.9& 1688& 11.48& 195.4981& 27.3878 &$72.4\pm3.7$&$50.8\pm2.6$&$21.8\pm1.7$&$29.4\pm2.0$&$14.7\pm 1.4$ \\
 38& 10.21&  --& 34.9& 1715& 7.02& 195.4899& 28.0056 &$21.8\pm3.3$&$15.3\pm2.3$&$11.9\pm2.0$&$ 3.4\pm1.3$&$ 0.0\pm 0.7$ \\
 45& 10.53& --& 47.7& 1750& 8.16& 195.4742& 27.6234  &$17.6\pm2.2$&$12.3\pm1.5$&$10.3\pm1.3$&$ 2.1\pm0.7$&$ 1.3\pm 0.6$\\
 63& 8.30& 2.60& 6.7& 1904& 8.51& 195.4243& 27.1215  &$ 6.5\pm1.9$&$ 4.6\pm1.3$&$ 2.3\pm1.0$&$ 2.3\pm0.9$&$ 0.4\pm 0.6$ \\
 71& 8.91& 2.27& 9.1& 1940& 10.40& 195.4088& 27.5915 &$ 2.1\pm1.1$&$ 1.5\pm0.8$&$ 0.5\pm0.6$&$ 1.0\pm0.6$&$-0.3\pm 0.3$  \\
 94& 10.78& --& 68.8& 2059& 5.72& 195.3586& 27.8860  &$22.9\pm2.5$&$16.1\pm1.7$&$10.0\pm1.3$&$ 6.1\pm1.1$&$ 0.1\pm 0.6$ \\
138& 10.63& --& 47.3& 2374& 7.41& 195.2339& 27.7908  &$94.1\pm4.0$&$66.0\pm2.8$&$40.2\pm2.2$&$26.1\pm1.8$&$13.3\pm 1.3$ \\
141& 10.40& --& 47.4& 2390& 8.97& 195.2269& 28.0064  &$18.4\pm2.6$&$12.9\pm1.8$&$ 4.7\pm1.2$&$ 8.2\pm1.4$&$ 2.7\pm 1.0$ \\
151& 10.17& --& 25.3& 2413& 5.93& 195.2165& 28.3653  &$ 6.9\pm1.3$&$ 4.9\pm0.9$&$ 4.4\pm0.8$&$ 0.6\pm0.5$&$ 0.0\pm 0.4$ \\
169& 10.06& --& 26.6& 2516& 14.12& 195.1740& 27.9754&$ 7.3\pm2.2$&$ 5.1\pm1.5$&$ 2.8\pm1.1$&$ 2.3\pm1.1$&$ 1.7\pm 0.8$  \\
177& 8.45& 1.36& 8.4& 2550& 2.87& 195.1602& 28.0098  &$ 7.9\pm2.0$&$ 5.6\pm1.4$&$ 3.3\pm1.0$&$ 2.3\pm1.0$&$ 1.8\pm 0.8$ \\
179& 10.02& --& 22.8& 2559& 15.45& 195.1598& 28.0576 &$16.1\pm2.1$&$11.3\pm1.5$&$ 8.7\pm1.1$&$ 2.5\pm1.0$&$ 0.4\pm 0.7$  \\
181& 8.21& 2.63& 4.9& 2566& 5.36& 195.1544& 28.2665  &$ 4.9\pm1.7$&$ 3.4\pm1.2$&$ 0.1\pm0.7$&$ 3.3\pm1.0$&$ 3.4\pm 0.9$ \\
184& 9.86& --& 20.7& 2599& 15.05& 195.1432& 27.6376  &$ 8.4\pm2.3$&$ 5.9\pm1.6$&$ 4.4\pm1.2$&$ 1.7\pm1.0$&$-0.7\pm 0.6$ \\
207& 9.12& 1.64& 10.4& 2725& 7.49& 195.0928& 28.3999 &$1163\pm11.$&$ 816\pm7.7$&$ 476\pm5.9$&$ 344\pm5.0$&$ 139\pm 3.2$  \\
209& 10.25& --& 26.7& 2798& 4.37& 195.0721& 27.9566  &$ 9.9\pm2.7$&$ 6.9\pm1.9$&$ 5.8\pm1.4$&$ 0.9\pm1.3$&$ 0.2\pm 1.0$ \\
225& 7.85& 1.26& 4.3& 2888& 5.89& 195.0433& 28.1300  &$16.4\pm2.0$&$11.5\pm1.4$&$ 6.8\pm1.0$&$ 4.8\pm1.0$&$ 2.3\pm 0.8$ \\
231& 11.14& --& 112.7& 2921& 8.51& 195.0343& 27.9766 &$55.4\pm4.1$&$38.9\pm2.8$&$17.8\pm2.0$&$21.4\pm2.1$&$ 4.3\pm 1.5$  \\
238& 10.26& --& 42.1& 2975& 7.14& 195.0180& 27.9878  &$11.5\pm2.9$&$ 8.1\pm2.0$&$ 4.6\pm1.4$&$ 3.5\pm1.5$&$ 1.1\pm 1.1$ \\
246& 10.30& --& 40.6& 3055& 5.26& 194.9904& 28.2467  &$ 7.3\pm2.3$&$ 5.2\pm1.6$&$ 3.6\pm1.2$&$ 1.7\pm1.1$&$ 2.0\pm 1.0$ \\
263& 7.83& 2.09& 4.9& 3152& 7.53& 194.9500& 27.9944  &$19.4\pm3.4$&$13.6\pm2.4$&$ 6.7\pm1.7$&$ 7.0\pm1.7$&$ 1.1\pm 1.3$ \\
265& 9.91& --& 24.8& 3170& 3.12& 194.9437& 27.9744   &$16.0\pm3.1$&$11.2\pm2.2$&$ 5.3\pm1.5$&$ 5.7\pm1.6$&$ 2.5\pm 1.2$\\
274& 9.48& --& 16.2& 3262& 19.11& 194.9186& 27.8589&$11.2\pm2.8$&$ 7.8\pm2.0$&$ 3.1\pm1.4$&$ 4.8\pm1.4$&$-0.1\pm 1.0$   \\
275& 9.74& --& 20.7& 3269& 11.21& 194.9161& 27.9544  &$ 5.4\pm3.0$&$ 3.8\pm2.1$&$ 2.1\pm1.5$&$ 1.5\pm1.5$&$ 1.2\pm 1.1$ \\
281& 11.08& --& 129.1& 3329& 5.24& 194.8985& 27.9589 &$45.5\pm3.2$&$31.9\pm2.2$&$18.8\pm1.6$&$13.2\pm1.6$&$ 1.5\pm 1.1$  \\
291& 9.44& --& 14.1& 3403& 15.10& 194.8797& 27.7900&$10.6\pm2.6$&$ 7.4\pm1.8$&$ 4.3\pm1.3$&$ 3.2\pm1.3$&$ 0.2\pm 0.9$   \\
294& 9.87& --& 20.2& 3423& 7.64& 194.8721& 27.8500   &$14.2\pm3.1$&$10.0\pm2.2$&$ 2.9\pm1.5$&$ 6.9\pm1.6$&$ 3.1\pm 1.2$\\
301& 10.38& --& 37.5& 3561& 26.35& 194.8329& 28.0911 &$ 6.7\pm1.8$&$ 4.7\pm1.3$&$ 2.6\pm0.9$&$ 2.1\pm0.9$&$ 0.7\pm 0.7$  \\
307& 9.28& 0.74& 16.1& 3585& 14.75& 194.8221& 27.5911&$ 6.6\pm2.4$&$ 4.7\pm1.7$&$ 2.2\pm1.2$&$ 2.4\pm1.2$&$ 0.3\pm 0.8$   \\
308& 8.38& 1.89& 6.2& 3606& 8.38& 194.8193& 27.8955  &$54.2\pm3.3$&$38.0\pm2.3$&$21.6\pm1.6$&$16.4\pm1.6$&$ 8.8\pm 1.2$ \\
312& 9.98& --& 36.1& 3656& 29.63& 194.8140& 28.0788  &$ 2.3\pm1.9$&$ 1.6\pm1.3$&$ 1.0\pm0.9$&$ 0.6\pm1.0$&$ 0.5\pm 0.7$ \\
314& 9.92& --& 20.9& 3661& 8.97& 194.8073& 27.4022   &$ 5.4\pm2.2$&$ 3.8\pm1.5$&$ 2.1\pm1.1$&$ 1.7\pm1.1$&$-0.2\pm 0.7$\\
315& 10.31& --& 36.1& 3664& 7.94& 194.8041& 27.9777  &$ 5.9\pm2.4$&$ 4.1\pm1.7$&$ 3.4\pm1.2$&$ 0.7\pm1.2$&$ 1.9\pm 0.9$ \\
324& 10.32& --& 38.6& 3792& 4.86& 194.7661& 28.1232  &$27.0\pm2.4$&$19.0\pm1.7$&$13.6\pm1.3$&$ 5.6\pm1.1$&$ 1.0\pm 0.8$ \\
326& 9.94& --& 23.3& 3816& 5.93& 194.7586& 28.1165   &$ 9.3\pm2.7$&$ 6.5\pm1.9$&$ 4.2\pm1.4$&$ 2.2\pm1.3$&$ 0.9\pm 0.9$\\
327& 10.02& --& 21.6& 3818& 7.86& 194.7571& 28.2254  &$ 4.4\pm2.0$&$ 3.1\pm1.4$&$ 2.7\pm1.0$&$ 0.3\pm0.9$&$ 1.2\pm 0.8$ \\
336& 10.14& --& 22.5& 3896& 11.62& 194.7314& 27.8321 &$12.0\pm2.8$&$ 8.4\pm2.0$&$ 7.0\pm1.5$&$ 1.4\pm1.3$&$-1.0\pm 1.0$  \\
347& 7.86& 2.05& 4.3& 4028& 5.93& 194.6936& 27.3565  &$25.9\pm2.6$&$18.2\pm1.8$&$11.1\pm1.4$&$ 7.0\pm1.2$&$ 2.4\pm 0.8$ \\
363& 9.95& --& 16.1& 4135& 7.14& 194.6540& 27.1753 &$13.7\pm3.0$&$ 9.6\pm2.1$&$ 3.4\pm1.3$&$ 6.2\pm1.6$&$ 0.3\pm 0.9$  \\
366& 10.44& --& 28.5& 4156& 6.28& 194.6466& 27.5964  &$18.8\pm2.4$&$13.2\pm1.7$&$10.2\pm1.3$&$ 3.0\pm1.0$&$-0.3\pm 0.6$ \\
369& 9.80& --& 18.6& 4159& 9.41& 194.6425& 27.2642   &$ 5.6\pm2.1$&$ 3.9\pm1.5$&$ 2.9\pm1.0$&$ 1.1\pm1.0$&$ 1.9\pm 0.8$\\
377& 9.00& 1.38& 13.9& 4188& 13.47& 194.6293& 28.3774&$ 4.1\pm1.4$&$ 2.9\pm1.0$&$ 2.2\pm0.8$&$ 0.7\pm0.6$&$-0.3\pm 0.4$   \\
382& 10.12& --& 29.2& 4230& 9.65& 194.6228& 28.0174  &$ 9.7\pm3.0$&$ 6.8\pm2.1$&$ 4.3\pm1.5$&$ 2.5\pm1.5$&$-1.4\pm 1.0$ \\
392& 10.04& --& 25.0& 4315& 11.07& 194.5915& 27.9662 &$11.3\pm3.2$&$ 7.9\pm2.2$&$ 6.6\pm1.7$&$ 1.3\pm1.5$&$ 2.0\pm 1.2$  \\
400& 9.45& --& 11.7& 4351& 10.05& 194.5774& 27.3095&$ 5.4\pm1.6$&$ 3.8\pm1.1$&$ 1.7\pm0.8$&$ 2.0\pm0.8$&$ 0.3\pm 0.5$   \\
401& 7.89& 0.91& 4.4& 4358& 5.27& 194.5702& 28.3061  &$ 8.0\pm2.1$&$ 5.6\pm1.4$&$ 2.5\pm1.0$&$ 3.3\pm1.0$&$ 1.6\pm 0.8$ \\
417& 10.39& --& 28.9& 4471& 6.63& 194.5237& 28.2427  &$34.0\pm2.5$&$23.8\pm1.8$&$17.6\pm1.5$&$ 6.6\pm1.0$&$ 0.4\pm 0.6$ \\
418& 9.75& --& 17.9& 4499& 13.02& 194.5166& 27.8149  &$ 4.0\pm1.3$&$ 2.8\pm0.9$&$ 1.6\pm0.7$&$ 1.2\pm0.7$&$ 0.5\pm 0.5$ \\
419& 8.90& 1.76& 6.1& 4513& 6.32& 194.5078& 28.4560  &$15.9\pm2.3$&$11.2\pm1.6$&$ 5.6\pm1.2$&$ 5.6\pm1.1$&$ 2.1\pm 0.8$ \\
425& 9.76& 1.50& 20.0& 4555& 5.76& 194.4905& 28.0604 &$15.6\pm2.0$&$10.9\pm1.4$&$ 7.1\pm1.1$&$ 4.0\pm0.9$&$ 1.1\pm 0.6$  \\
432& 9.64& --& 20.7& 4592& 10.17& 194.4785& 27.6148  &$ 5.2\pm1.2$&$ 3.7\pm0.9$&$ 2.9\pm0.7$&$ 0.9\pm0.6$&$ 0.9\pm 0.5$ \\
433& 9.50& 1.11& 16.1& 4579& 16.49& 194.4774& 27.5814&$ 4.4\pm1.1$&$ 3.1\pm0.8$&$ 1.4\pm0.6$&$ 1.8\pm0.6$&$-0.2\pm 0.3$   \\
434& 9.64& --& 19.5& 4597& 9.59& 194.4766& 27.4903   &$ 3.8\pm1.3$&$ 2.7\pm0.9$&$ 1.9\pm0.7$&$ 0.8\pm0.6$&$ 0.5\pm 0.5$\\
438& 7.99& 1.80& 4.2& 4615& 6.82& 194.4669& 27.6981  &$ 3.5\pm1.1$&$ 2.5\pm0.8$&$ 1.3\pm0.6$&$ 1.2\pm0.6$&$ 0.0\pm 0.4$ \\
444& 9.80& --& 17.1& 4648& 19.21& 194.4509& 28.1747  &$ 3.3\pm1.2$&$ 2.3\pm0.9$&$ 2.0\pm0.7$&$ 0.4\pm0.6$&$ 0.1\pm 0.4$ \\
454& 8.71& 2.27& 8.1& 4715& 7.41& 194.4297& 27.6068  &$ 6.5\pm1.1$&$ 4.5\pm0.8$&$ 0.8\pm0.5$&$ 3.7\pm0.6$&$ 7.9\pm 0.7$ \\
456& 8.88& 1.79& 10.2& 4718& 8.10& 194.4238& 27.7957 &$21.5\pm1.5$&$15.0\pm1.1$&$ 7.0\pm0.7$&$ 8.1\pm0.8$&$ 5.1\pm 0.6$  \\
464& 9.67& --& 14.8& 4792& 8.27& 194.4002& 27.4856 &$ 3.2\pm1.5$&$ 2.2\pm1.0$&$ 1.3\pm0.7$&$ 0.9\pm0.7$&$ 1.1\pm 0.6$  \\
466& 10.09& --& 23.9& 4794& 7.01& 194.3989& 27.4934  &$ 4.1\pm1.3$&$ 2.9\pm0.9$&$ 2.8\pm0.7$&$ 0.1\pm0.6$&$ 0.0\pm 0.4$ \\
473& 10.25& --& 32.6& 4829& 7.36& 194.3870& 27.6100  &$15.3\pm1.4$&$10.7\pm1.0$&$ 8.4\pm0.8$&$ 2.4\pm0.6$&$ 1.1\pm 0.4$ \\
476& 7.88& 1.95& 4.5& 4843& 3.56& 194.3819& 27.3867  &$17.3\pm2.3$&$12.1\pm1.6$&$ 4.1\pm1.0$&$ 7.9\pm1.2$&$ 3.5\pm 0.9$ \\
477& 9.87& --& 20.8& 4849& 12.87& 194.3765& 28.1888  &$ 3.1\pm1.1$&$ 2.1\pm0.7$&$ 1.5\pm0.6$&$ 0.7\pm0.5$&$ 0.4\pm 0.4$ \\
484& 9.82& --& 15.1& 4907& 11.02& 194.3586& 27.5443  &$ 0.5\pm1.0$&$ 0.3\pm0.7$&$-0.3\pm0.5$&$ 0.6\pm0.5$&$ 0.1\pm 0.4$ \\
486& 9.78& --& 15.8& 4918& 7.98& 194.3543& 27.4032 &$ 7.2\pm2.0$&$ 5.0\pm1.4$&$ 4.7\pm1.1$&$ 0.3\pm0.8$&$ 1.1\pm 0.7$  \\
489& 10.79& --& 79.5& 4928& 8.34& 194.3513& 27.4976  &$71.6\pm3.3$&$50.2\pm2.3$&$29.3\pm1.8$&$21.4\pm1.5$&$ 4.9\pm 0.9$ \\
498& 9.51& --& 10.3& 4987& 4.83& 194.3193& 27.6186 &$ 2.7\pm1.0$&$ 1.9\pm0.7$&$ 1.2\pm0.5$&$ 0.7\pm0.5$&$ 0.5\pm 0.3$  \\
515& 9.74& --& 19.9& 5038& 5.59& 194.2942& 27.4041   &$65.2\pm3.1$&$45.7\pm2.1$&$27.7\pm1.6$&$18.3\pm1.4$&$ 9.1\pm 1.0$\\
558& 9.11& 1.81& 16.5& 5254& 2.14& 194.1949& 27.2936 &$ 9.2\pm1.7$&$ 6.4\pm1.2$&$ 3.8\pm0.9$&$ 2.6\pm0.8$&$ 1.0\pm 0.5$  \\
560& 10.52& --& 49.7& 5279& 11.27& 194.1820& 27.1780 &$33.9\pm3.1$&$23.8\pm2.2$&$20.3\pm2.0$&$ 3.5\pm1.0$&$ 0.9\pm 0.6$  \\
569& 9.81& --& 20.1& 5364& 8.13& 194.1418& 27.5377   &$ 1.8\pm1.4$&$ 1.3\pm1.0$&$ 1.0\pm0.7$&$ 0.4\pm0.7$&$-0.2\pm 0.5$\\
\hline
\end{tabular}
}
\end{center}
}
\end{table*}

\begin{table*}[ht]
{
\begin{center}
\footnotesize
{\renewcommand{\arraystretch}{0.6}\renewcommand{\tabcolsep}{0.12cm}
\caption{\footnotesize
Association of entropy fluctuations in the Coma centre with galaxies.
\label{t:s}}

\begin{tabular}{cccccccccccccc}
 \hline
S  & $lg(L_b)$ & B-R &  $r_e$ & GMP & separ.& \multicolumn{2}{c}{X-ray source J2000} & $F_{0.5-2}$ $10^{-15}$ & \multicolumn{4}{c}{$10^{-3}$ counts s$^{-1}$, band (keV):}\\
numb. & $L_\odot$ & &$^{\prime\prime}$ & & $^{\prime\prime}$&\multicolumn{2}{c}{RA Decl.} &ergs s$^{-1}$ cm$^{-2}$&0.5--2&0.5--1 & 1--2& 2-4.5\\
\hline
  6& 10.63& --& 47.3& 2374& 11.63& 195.2351& 27.7908&$92.4\pm4.0$ &$64.8\pm2.8$ &$39.1\pm2.1$ &$26.0\pm1.8$ &$12.7\pm1.3$\\
  7& 10.40& --& 47.4& 2390&  8.97& 195.2269& 28.0064&$11.5\pm2.4$ &$ 8.1\pm1.7$ &$ 2.1\pm1.1$ &$ 5.9\pm1.3$ &$ 1.4\pm0.9$\\
 22& 10.06& --& 26.6& 2516& 14.20& 195.1727& 27.9698&$ 3.6\pm2.0$ &$ 2.5\pm1.4$ &$ 1.9\pm1.0$ &$ 0.6\pm1.0$ &$ 0.9\pm0.8$\\
 26& 10.02& --& 22.8& 2559& 17.12& 195.1590& 28.0565&$15.4\pm2.0$ &$10.8\pm1.4$ &$ 8.8\pm1.1$ &$ 1.9\pm0.9$ &$ 0.3\pm0.7$\\
 43&  8.30&1.9&  7.7& 2755&  8.57& 195.0797& 27.9944&$ 3.4\pm2.4$ &$ 2.4\pm1.7$ &$ 3.0\pm1.2$ &$-0.6\pm1.2$ &$-0.4\pm0.9$\\
 45&  9.72& --& 18.0& 2798&  1.80& 195.0733& 27.9566&$10.6\pm2.6$ &$ 7.4\pm1.8$ &$ 6.2\pm1.3$ &$ 1.1\pm1.3$ &$ 0.3\pm1.0$\\
 50&  8.90&1.4& 13.3& 2856& 16.24& 195.0495& 28.0533&$ 8.9\pm2.1$ &$ 6.2\pm1.4$ &$ 2.9\pm1.0$ &$ 3.2\pm1.0$ &$-0.2\pm0.7$\\
 52&  7.85&1.3&  4.3& 2888&  5.89& 195.0433& 28.1300&$15.8\pm2.0$ &$11.1\pm1.4$ &$ 6.6\pm1.0$ &$ 4.6\pm1.0$ &$ 2.2\pm0.7$\\
 54&  9.40& --& 12.7& 2897& 12.34& 195.0418& 27.8611&$ 3.6\pm2.6$ &$ 2.5\pm1.8$ &$ 0.9\pm1.2$ &$ 1.7\pm1.3$ &$-0.8\pm0.9$\\
 56& 11.14& --&112.7& 2921&  8.82& 195.0343& 27.9777&$50.9\pm3.7$ &$35.7\pm2.6$ &$17.6\pm1.8$ &$18.4\pm1.9$ &$ 3.4\pm1.3$\\
 59&  9.48& --& 17.8& 2960& 10.25& 195.0205& 28.0222&$10.4\pm2.4$ &$ 7.3\pm1.6$ &$ 4.4\pm1.2$ &$ 2.8\pm1.2$ &$ 2.9\pm0.9$\\
 60& 10.26& --& 42.1& 2975&  2.73& 195.0167& 27.9878&$10.7\pm2.8$ &$ 7.5\pm1.9$ &$ 3.8\pm1.4$ &$ 3.8\pm1.4$ &$ 1.1\pm1.0$\\
 76&  7.87&1.2&  5.6& 3107&  5.24& 194.9651& 27.9956&$14.3\pm2.6$ &$10.0\pm1.9$ &$ 4.6\pm1.3$ &$ 5.7\pm1.3$ &$ 2.0\pm1.0$\\
 80&  7.83&2.1&  4.9& 3152&  7.74& 194.9500& 27.9933&$16.6\pm2.6$ &$11.7\pm1.8$ &$ 6.6\pm1.3$ &$ 5.1\pm1.3$ &$ 1.0\pm0.9$\\
 82&  8.46&1.2&  9.2& 3220& 10.39& 194.9261& 28.1333&$ 1.8\pm1.7$ &$ 1.2\pm1.2$ &$ 0.2\pm0.8$ &$ 1.0\pm0.9$ &$-0.2\pm0.6$\\
 89& 11.08& --&129.1& 3329&  5.24& 194.8985& 27.9589&$37.4\pm3.1$ &$26.2\pm2.2$ &$16.1\pm1.5$ &$10.2\pm1.5$ &$-0.2\pm1.1$\\
 95& 10.13& --& 30.5& 3367&  3.29& 194.8846& 27.9833&$ 4.8\pm2.8$ &$ 3.4\pm1.9$ &$ 1.4\pm1.3$ &$ 2.0\pm1.4$ &$-0.1\pm1.0$\\
 98&  9.44& --& 14.1& 3403& 15.27& 194.8759& 27.7878&$ 4.0\pm2.4$ &$ 2.8\pm1.7$ &$ 2.1\pm1.2$ &$ 0.9\pm1.2$ &$ 0.3\pm0.9$\\
 99&  9.87& --& 20.2& 3423& 12.14& 194.8734& 27.8500&$15.9\pm3.0$ &$11.2\pm2.1$ &$ 4.2\pm1.5$ &$ 6.9\pm1.5$ &$ 3.1\pm1.1$\\
100&  9.61& --& 22.8& 3471& 11.86& 194.8582& 28.0022&$ 1.8\pm2.3$ &$ 1.2\pm1.6$ &$ 1.1\pm1.1$ &$ 0.3\pm1.2$ &$-0.9\pm0.8$\\
102& 10.20& --& 34.0& 3510& 32.71& 194.8545& 27.9077&$ 1.4\pm2.7$ &$ 1.0\pm1.9$ &$ 0.1\pm1.3$ &$ 0.9\pm1.3$ &$-1.0\pm1.0$\\
107& 10.38& --& 37.5& 3561& 26.35& 194.8329& 28.0911&$ 3.9\pm1.7$ &$ 2.7\pm1.2$ &$ 1.8\pm0.8$ &$ 0.9\pm0.9$ &$ 0.2\pm0.6$\\
113& 10.31& --& 36.1& 3664& 16.78& 194.8066& 27.9777&$ 1.5\pm2.4$ &$ 1.1\pm1.7$ &$ 1.2\pm1.2$ &$-0.2\pm1.2$ &$ 0.2\pm0.9$\\
115&  8.30&2.5&  7.7& 3702&  9.62& 194.7916& 27.8866&$ 0.0\pm2.5$ &$-0.4\pm1.7$ &$ 1.3\pm1.2$ &$-1.7\pm1.2$ &$ 0.3\pm0.9$\\
119&  9.96& --& 26.1& 3761& 22.91& 194.7751& 27.9910&$ 1.2\pm2.2$ &$ 0.8\pm1.5$ &$-1.0\pm1.0$ &$ 2.0\pm1.1$ &$-0.3\pm0.8$\\
120& 10.08& --& 32.9& 3730& 27.02& 194.7755& 27.7877&$ 0.1\pm2.2$ &$ 0.1\pm1.5$ &$-0.6\pm1.1$ &$ 0.7\pm1.1$ &$ 0.4\pm0.9$\\
123& 10.32& --& 38.6& 3792&  6.51& 194.7661& 28.1254&$26.5\pm2.3$ &$18.6\pm1.6$ &$12.8\pm1.2$ &$ 6.0\pm1.1$ &$ 0.6\pm0.7$\\
124&  9.94& --& 23.3& 3816& 22.35& 194.7560& 28.1099&$ 8.6\pm2.0$ &$ 6.1\pm1.4$ &$ 4.5\pm1.0$ &$ 1.5\pm1.0$ &$ 1.3\pm0.8$\\
130& 10.14& --& 22.5& 3896& 11.62& 194.7314& 27.8321&$12.1\pm2.7$ &$ 8.5\pm1.9$ &$ 6.9\pm1.4$ &$ 1.5\pm1.3$ &$-0.9\pm0.9$\\
131&  9.42& --& 13.8& 3943& 12.43& 194.7214& 27.8165&$ 7.2\pm2.6$ &$ 5.0\pm1.8$ &$ 1.8\pm1.2$ &$ 3.1\pm1.3$ &$-1.2\pm0.9$\\
141&  8.19&1.7&  5.4& 4148&  2.06& 194.6454& 28.0486&$ 1.4\pm2.6$ &$ 1.0\pm1.8$ &$ 0.1\pm1.3$ &$ 0.8\pm1.3$ &$-0.4\pm0.9$\\
147& 10.12& --& 29.2& 4230& 15.42& 194.6215& 28.0185&$ 8.1\pm2.9$ &$ 5.7\pm2.1$ &$ 2.4\pm1.4$ &$ 3.5\pm1.5$ &$-0.7\pm1.0$\\
149& 10.04& --& 25.0& 4315& 11.07& 194.5915& 27.9662&$11.0\pm3.0$ &$ 7.7\pm2.1$ &$ 6.5\pm1.6$ &$ 1.1\pm1.4$ &$ 1.3\pm1.1$\\
\hline                                                   
\end{tabular}                                            
%
%
%
}
\end{center}
}
\end{table*}

\begin{table*}[ht]
\begin{center}
\footnotesize
\renewcommand{\arraystretch}{0.9}\renewcommand{\tabcolsep}{0.12cm}
\caption{\footnotesize
X-ray emission of the Coma cluster galaxies, with membership verified using
the Goldmine database.
\label{t:gm}}

\begin{tabular}{cccccccccccc}
 \hline
ID & $lg(L_b)$ &   $r_e$ $^{\prime\prime}$ & CGCG &     $V_h$  &   U  &   B &   V & GMP   \\
\hline
X019&  9.99& 15.9& 160108& 8323& 15.72& 15.86& 15.34& 1576 \\
X025& 10.15& 22.4& 160106& 6876& 15.43& 15.34& 14.66& 1616 \\
X033& 10.03& 12.8& 160104& 7213&  0.00& 15.79&  0.00& 1681 \\
X038& 10.27& 34.9& 160105& 7747&  0.00& 14.90& 13.81& 1715 \\
X045& 10.55& 47.7& 160103& 7874&  0.00& 14.24& 13.17& 1750 \\
X094& 10.71& 68.8& 160095& 5482& 14.03& 13.52& 12.55& 2059 \\
X138& 10.71& 47.3& 160260& 7985& 14.01& 13.73& 12.86& 2374 \\
X141& 10.27& 47.4& 160259& 4964& 15.06& 14.52& 13.52& 2390 \\
X151& 10.23& 25.3& 160091& 7647&  0.00& 15.06&  0.00& 2413 \\
X179& 10.15& 22.8& 160252& 7718& 15.19& 15.38& 14.71& 2559 \\
X184& 10.03& 20.7& 160086& 7499& 15.61& 15.77& 15.36& 2599 \\
X209& 10.31& 26.7& 160248& 6848& 15.29& 14.79& 13.76& 2798 \\
X231& 10.99&112.7& 160241& 6517& 13.18& 12.59& 11.55& 2921 \\
X238& 10.15& 42.1& 160239& 6397& 15.71& 15.36& 14.36& 2975 \\
X246& 10.31& 40.6& 160238& 6730& 15.29& 14.77& 13.74& 3055 \\
X265&  9.95& 24.8& 160236& 9400& 16.33& 15.87& 14.84& 3170 \\
X281& 10.71&129.1& 160231& 7189& 13.68& 13.12& 12.09& 3329 \\
X294&  9.99& 20.2& 160226& 6950& 16.59& 15.82& 14.79& 3423 \\
X301& 10.35& 37.5& 160224& 4643& 15.26& 14.71& 13.74& 3561 \\
X314& 10.03& 20.9& 160074& 5633&  0.00& 15.68&  0.00& 3661 \\
X315& 10.27& 36.1& 160221& 6760& 15.19& 14.60& 13.55& 3664 \\
X324& 10.31& 38.6& 160215& 7966& 15.28& 14.74& 13.69& 3792 \\
X326&  9.99& 23.3& 160213& 9386& 15.61& 15.85& 15.26& 3816 \\
X327& 10.07& 21.6& 160214& 8028&  0.00& 15.45&  0.00& 3818 \\
X336& 10.23& 22.5& 160212& 7549& 15.38& 15.10& 14.30& 3896 \\
X363& 10.03& 16.1& 160067& 7653& 15.40& 15.58& 15.19& 4135 \\
X366& 10.51& 28.5& 160068& 7550& 14.65& 14.44& 13.67& 4156 \\
X369& 10.03& 18.6& 160064& 7368&  0.00& 15.98& 15.59& 4159 \\
X382& 10.19& 29.2& 160065& 7188&  0.00& 15.17&  0.00& 4230 \\
X392& 10.07& 25.0& 160063& 6044& 15.92& 15.36& 14.36& 4315 \\
X417& 10.51& 28.9& 160055& 7164& 14.53& 14.46& 13.78& 4471 \\
X418&  9.91& 17.9& 160053& 7205&  0.00& 16.04&  0.00& 4499 \\
X444&  9.99& 17.1& 160049& 7245&  0.00& 15.96& 14.98& 4648 \\
X466& 10.23& 23.9& 160046& 7317&  0.00& 15.32&  0.00& 4794 \\
X473& 10.27& 32.6& 160042& 6097& 15.45& 14.88& 13.86& 4829 \\
X477& 10.03& 20.8& 160043& 7078&  0.00& 15.61&  0.00& 4849 \\
X484& 10.07& 15.1& 160040& 5523&  0.00& 15.90&  0.00& 4907 \\
X489& 10.75& 79.5& 160039& 7318& 13.81& 13.29& 12.21& 4928 \\
X560& 10.55& 49.7& 160028& 7610&  0.00& 14.21& 13.20& 5279 \\
S119&  9.95& 26.1& 160216& 7895& 16.12& 15.62& 14.59& 3761 \\
\hline 
\end{tabular}
\end{center}
\end{table*}

\begin{table*}[ht]
\begin{center}
\renewcommand{\arraystretch}{0.9}\renewcommand{\tabcolsep}{0.12cm}
\caption{\footnotesize
Additional list of X-ray detected galaxies with identified membership to
the Coma cluster using sources other than Goldmine.
\label{t:gm2}}

\begin{tabular}{cccccc}
 \hline
ID & $lg(L_b)$ &   $r_e$ $^{\prime\prime}$  & GMP & memb.&$V_h$/comment  \\
\hline
X169&10.06&  26.6& 2516& M & 6363\\
S054& 9.40&12.7& 2897&M?&9902/confusion\\
S059& 9.48&17.8& 2960&M&5786\\
S095&10.13&30.5& 3367&M& 5848\\
X291& 9.44&  14.1& 3403& M &7825\\
S100& 9.61&22.8& 3471&M&6684\\
S131& 9.42&13.8& 3943&M&5496\\
X400&9.45&11.7& 4351&M & 7447\\
X425& 9.76&  20.0& 4555& M&8299 \\
X433&9.50&16.1& 4579&M &4999\\
X464&9.67&14.8& 4792&M&7234\\
X486&9.78& 15.8& 4918&M&4857\\
X498&9.51&10.3& 4987&M&7257 \\
X515&9.74&19.9& 5038&M&6215\\
X558&9.11&16.5& 5254&M&7787\\
X569&9.81&20.1&5364&M&7105\\
X177&8.45&8.4&2550&M?&merger\\
X456&8.88&10.2&4718&M?& FOCA\\
X476&7.88&4.5&4843&B?&\\
X181&8.21& 4.9& 2566&  B&\\
X207&9.12&10.4& 2725&  B&27281\\
X274&9.48&16.2& 3262&  F& 3747 \\
X419&8.90& 6.1& 4513&  B&\\
X432&9.64&20.7& 4592&B & 20320\\
S141&8.19& 5.4& 4148&B &\\
\hline 
\end{tabular}
\end{center}
\end{table*}

\end{document}